\documentclass[journal]{IEEEtran}
\usepackage{xcolor}
\usepackage{graphicx}
\usepackage{tikz}
\usepackage{color}
\usepackage{amsmath}
\usepackage{subfig}
\usepackage{algorithm}
\usepackage{algorithmic}
\usepackage{amssymb}
\graphicspath{{./Fig/}}
\usepackage{math}
\usepackage{pifont}

\usepackage[final]{review}
\setcoverletter{coverletter-R1.tex}

\begin{document}
\bibliographystyle{ieeetr}
\title{Multiple-Speaker Localization Based on Direct-Path Features and Likelihood Maximization with Spatial Sparsity Regularization
}
 \author{Xiaofei Li,
         Laurent~Girin,
         Radu Horaud and
         Sharon Gannot
 \thanks{X. Li and R. Horaud are with INRIA Grenoble Rh\^one-Alpes, Montbonnot Saint-Martin, France. 
 E-mail: \texttt{first.last@inria.fr}
 }
 \thanks{L. Girin is with INRIA Grenoble Rh\^one-Alpes and with Univ. Grenoble Alpes, GIPSA-lab, Grenoble, France. 
 E-mail: \texttt{laurent.girin@gipsa-lab.grenoble-inp.fr}
 }%
 \thanks{Sharon Gannot is with Bar Ilan University, Faculty of Engineering, Israel. 
 E-mail: \texttt{Sharon.Gannot@biu.ac.il}
 }
 \thanks{This work was supported by the EU FP7 STREP project EARS \#609465 and by 
 the ERC Advanced Grant VHIA \#340113.}
 }

\maketitle

%

\begin{abstract}
This paper addresses the problem of multiple-speaker localization in noisy and reverberant environments, using binaural recordings of an acoustic scene.
A complex-valued Gaussian mixture model (CGMM) is adopted, whose components correspond to all the possible candidate source locations defined on a grid. 
After optimizing the CGMM-based objective function, given an observed set of complex-valued binaural features, both the number of sources and their locations are estimated by selecting the CGMM components with the largest weights. This is achieved by enforcing  a sparse solution, thus favoring a small number of speakers with respect to the large number of initial candidate source locations.
An entropy-based penalty term is added to the likelihood, thus imposing sparsity over the set of CGMM component weights.
In addition, the direct-path relative transfer function (DP-RTF) is used to build robust binaural features. The DP-RTF, recently proposed for single-source localization, was shown to be robust to reverberations, since it encodes inter-channel information corresponding to the direct-path of sound propagation. 
In this paper, we extend the DP-RTF estimation to the case of multiple sources. 
In the short-time Fourier transform domain, a consistency test is proposed to check whether a set of consecutive frames is associated to the same source or not. Reliable DP-RTF features are selected from the frames that pass the consistency test to be used for source localization. Experiments carried out using both simulation data and real data recorded with a robotic head confirm the efficiency of the proposed multi-source localization method. 
\end{abstract}


\section{Introduction}
\label{sec:introduction}

Multiple-speaker localization is an auditory scene analysis module with many applications in human-computer and human-robot interaction, video conferencing, etc. In this paper we address the multiple-speaker localization problem in the presence of noise and in reverberant environments. While we use binaural recordings of the acoustic scene, the method can be easily generalized to an arbitrary number of microphones.

Whenever there are more sources than microphones, which is the case in the present work, the so-called W-disjoint orthogonality (WDO) of the speech sources \cite{rickard2002,yilmaz2004} is widely employed by multiple-speaker localization methods.
The principle is that in each small region of the time-frequency (TF) domain, the audio signal is assumed to be dominated by only one source, because of the natural sparsity of speech signals in this domain. Therefore, multiple-speaker localization from binaural recordings can be decomposed in the following three-step process: (i)~binaural TF-domain localization features are extracted from the binaural signals using the short-time Fourier transform (STFT), or another TF decomposition; (ii)~these features are clustered into sources, and (iii) the clustered features are mapped to the source locations. 


Traditionally, the binaural features used for localization are the interaural level difference (ILD) and interaural time (or phase) difference (ITD or IPD), e.g., \cite{yilmaz2004,mandel2010,raspaud2010,may2011,traa2014}. Complex-valued features can also be used \cite{araki2007,winter2007,arberet2010}, as well as the relative phase ratio \cite{schwartz2014,dorfan2015}, since they can be easily clustered.  
However, these features are not robust to the presence of noise and reverberations. To reduce the noise effects, unbiased relative transfer function (RTF) estimators were adopted, such as the ones based on noise stationarity versus the non-stationarity of the desired signal \cite{gannot2001,dvorkind2005,mine2015assp}, 
or on the probability of speech-presence and spectral subtraction \cite{mine2015assp,cohen2004}, or on complex t-distribution \cite{deleforge2016rectified}. The RTF estimation is generalized to multiple sources in \cite{deleforge2015towards}.
To robustly estimate localization features in the presence of reverberations, the precedence effect \cite{litovsky1999} can be exploited, relying on the principle that signal onsets are dominated by the direct path. 
Interaural coherence \cite{faller2004}, coherence test \cite{mohan2008} and direct-path dominance test \cite{nadiri2014} were proposed to detect the frames dominated by one active source, from which localization features that are robust to reverberations can be estimated. However, in practice, significant reflection components often remain in the frames selected by these methods, due to an inaccurate model or to an improper decision threshold. 
In \cite{woodruff2012}, the TF bins dominated by one same source are grouped together based on the use of monaural features (such as pitch and onset/offset). 

To localize multiple active speakers using binaural features, many models have been developed. The simplest one, assuming free-field recording with small inter-microphone distance and low reverberations, rely on frequency-independent ITD features.   
Histogram methods \cite{yilmaz2004,faller2004} and k-means clustering \cite{araki2007} were then proposed to group these features and localize/separate the sources. When the inter-microphone distance is larger, the problem becomes more complex since the features derived from phase measures (IPD and ITD) are generally ambiguous along frequency due to phase wrapping. 
In \cite{mandel2010,raspaud2010}, the ITD ambiguity along frequency is solved by jointly exploiting the ILD. 
Frequency-wise clustering can be adopted, such as hierarchical clustering \cite{winter2007} and weighted sequential clustering \cite{arberet2010}. 
The frequency-wise clustering  faces the so-called source permutation problem, i.e. the indexing of clusters can be different from one frequency to the other.
To solve this problem, the speech spectrum correlation between frequencies is exploited in \cite{arberet2010}. A maximum likelihood method is proposed to formulate the source localization problem in \cite{schwartz2016}.
Based on manifold learning, two semi-supervised localization methods are proposed in \cite{laufer2016,laufer2017}.
A probabilistic mixture of linear regressions is used in \cite{deleforge2014mapping} to map a high-dimensional binaural feature vector (concatenated across frequencies) onto source location. 
In \cite{deleforge2014mapping}, only one source is considered. In \cite{deleforge2015acoustic} the method is extended to multiple sources relying on the WDO assumption. In \cite{deleforge2015colocalization} it is also extended to the direct colocalization of two sources without relying on the WDO assumption and source clustering.   

More often, solving the feature ambiguity and/or source permutation problems amounts to ensure the continuity of binaural cues (in particular IPD) across frequencies. 
IPD profiles as a function of frequency can be unwrapped using the direct-path propagation model of potential source locations. 
In \cite{sawada2004,sawada2007}, permutation alignment is processed by minimizing the cost function between the observations and the propagation model. 
In \cite{woodruff2012}, the azimuth set that has the largest likelihood given the feature observations is exhaustively searched from all the potential azimuth sets.
In \cite{may2011} a Gaussian mixture model (GMM) is used to learn the azimuth-dependent ambiguous ITD space of candidate sources, and the most likely azimuth with respect to the observed ITDs is estimated as the source direction.
Probabilistic models, mostly GMMs, were also proposed to both cluster and map the features onto source location \cite{mandel2010,traa2014,schwartz2014}. 
Here, source localization amounts to estimating the mixture model parameters from measured features, and then detect the main mixture components. 
In \cite{traa2014} a mixture of warped lines is fitted to the IPD observation profiles. Each warped line corresponds to a source direction. 
In \cite{mandel2010} each candidate interchannel time delay is considered as a GMM component. A mixture of GMMs is constructed to represent multiple sources. 
The azimuth of each source is given by the component that has the highest weight in the corresponding GMM.
A similar approach is proposed in \cite{schwartz2014}, but with GMM components corresponding to candidate 2D source positions thanks to the use of several pairs of microphones.

Recently, a probabilistic clustering method was proposed in \cite{dorfan2015} to localize an unknown number of emitting speech sources hypothetically located on a regular grid (as the method in \cite{malioutov2005}), 
where each grid-point location is known with respect to several microphone pairs. 
The relative phase ratio (RPR) associated with a microphone pair is predicted from the propagation model for each grid point and for each frequency. 
A set of complex-valued Gaussian mixture models (CGMMs), one mixture model per frequency, is built such that the number of components equals the number of grid points, the mixture components are centered around the predicted RPRs and they share the same fixed variance. Note that unlike \cite{mandel2010,schwartz2014} that use a separate GMM for each source, a common CGMM is used for all sources in \cite{dorfan2015}.
Since the mixture means  (predicted RPRs) and the variances are fixed, only the mixture weights have to be estimated. 
\addnote[dorfan]{1}{Notably, these weights are shared by all the mixtures over the frequency bins.
An EM algorithm alternates between assigning RPR observations to the mixture components (expectation) and estimating the weights (maximization). At convergence, the algorithm yields a weight value for each grid point and the number and location of active sources is obtained by applying a threshold to these weight values. Note that having one common source location candidate per mixture component shared across frequencies avoid the source permutation problem mentioned above.}
 
 \begin{figure*}[t]
\centering
\vspace{-2mm}
{\includegraphics[width=1\textwidth]{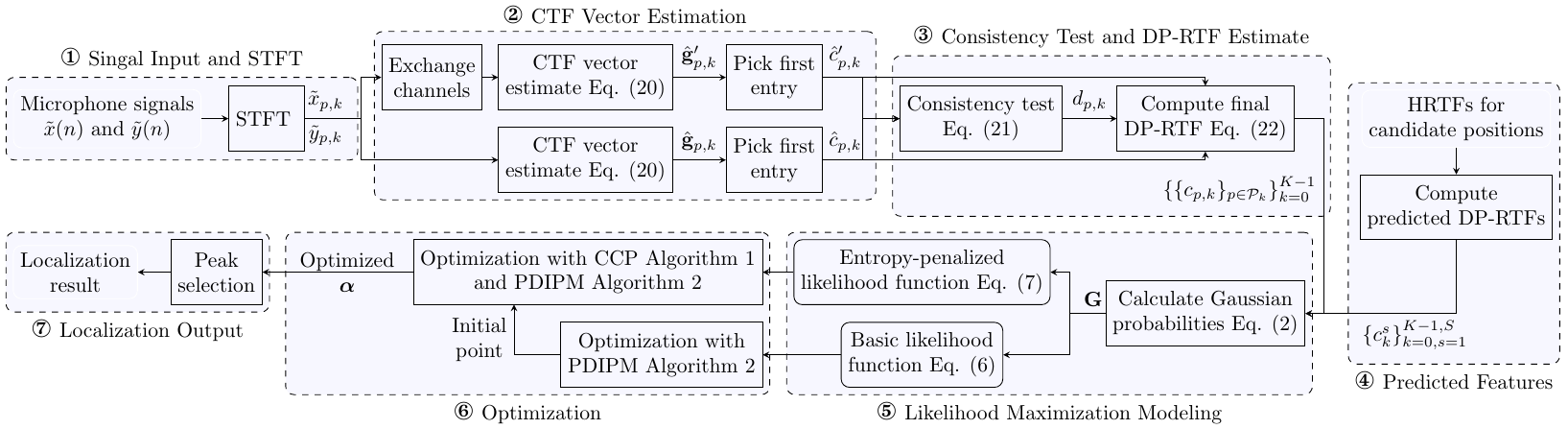}}
\caption{\small{Flowchart of the proposed sound source localization method.}} 
\label{figfc}
\vspace{-2mm}
\end{figure*}

\addnote[dirichlet]{1}{
The idea of placing many Gaussian components on a regular grid and of selecting a small number of components corresponding to the true audio sources has some similarities with sparse finite mixture modeling, namely to deliberately specify an overfitting mixture model with too many components \cite{figueiredo2002unsupervised,rousseau2011asymptotic,malsiner2016model}. In this setting, sparse solutions in terms of the number of components have been proposed in a Bayesian formulation within either variational inference \cite{bishop2007pattern} or sampling strategies \cite{malsiner2016model}. To obtain a sparse solution an appropriate prior on the weight distribution must be selected and a popular choice is the Dirichlet distribution \cite{ishwaran2001bayesian,bishop2007pattern,malsiner2016model}. The choice of the hyper-parameter of this distribution must guarantee that superfluous components are emptied automatically.
The rigorous asymptotic analysis of \cite{rousseau2011asymptotic} suggests that the Dirichlet hyper-parameter should be smaller than half the dimension of the parameter vector characterizing a mixture component. This theoretical result holds as the number of observations goes to infinity. In the case of a finite number of observations, it is necessary to select a much smaller value for the hyper-parameter \cite{bishop2007pattern,malsiner2016model}. While the use of Dirichlet priors is appealing from a Bayesian perspective, in practice there are some problems. First, it is not clear how to learn from the data how much sparsity is needed, i.e. how to choose the hyper-parameter. Second, one has to remove the emptied components by checking for small weights, which amounts to thresholding the Dirichlet posterior distribution. Third, we note that in our case the means are constrained by the acoustic model, therefore a full Bayesian treatment may not be justified.
}
 
The spatial sparsity is investigated in \cite{malioutov2005} for multi-source localization in a source signals reconstruction framework. In \cite{malioutov2005} the multiple sources are hypothetically located on a regular grid. The mixing matrix is known, and is composed of the steering vectors to all grid points. In anechoic environment, the steering vectors are given by the free-field sound propagation model. 
The signal reconstruction is formulated as an $\ell_2$ fit between the received signals and the source images. Since only a few grid points correspond to active sources, an $\ell_1$ regularization is used to impose the spatial sparsity of the source signals. 
In \cite{asaei2014}, the sparse signal reconstruction problem is extended to the reverberant environment. 
 
In this paper a new multiple-speaker localization method is proposed, based on the CGMM of \cite{dorfan2015} and associated likelihood function, combined with the use of direct-path relative transfer function (DP-RTF) as a binaural feature \cite{mine2016ITASL}. \addnote[flowchart_explanation]{1}{A block diagram of the proposed method is given in Fig.~\ref{figfc}. } This paper has the following contributions.  
First, it is proposed to minimize the negative log-likelihood function by adding an entropy-based penalty (Block \ding{176} in Fig.~\ref{figfc}), which enforces a sparse solution in terms of the free model parameters, i.e. the component weights. 
This corresponds to enforcing the spatial sparsity of sources, i.e. selecting a small number of active sources among the large number of potential source locations, in the spirit of \cite{malioutov2005} but implemented in a very different manner. 
Second, it is shown that the minimization of this penalized objective function can be carried out via a convex-concave optimization procedure (CCP) \cite{yuille2003} which is implemented using \cite{smola2005}: at each iteration, the concave penalty is approximated by its first-order Taylor expansion, such that the convex-concave problem becomes convex. 
The latter is solved using the primal-dual interior point method (PDIPM) \cite{boyd2004} (Block \ding{177} in Fig.~\ref{figfc}). 
Thirdly, it is proposed to use DP-RTF binaural features instead of RPR features. 
The DP-RTF is defined as the ratio between the direct path of the acoustic transfer function of the left and right channels. Unlike RPR, RTF and similar features, which are polluted by reverberations, the DP-RTF bears mainly the desired localization information. 
\addnote[ctf]{1}{The DP-RTF is estimated based on the convolutive transfer function (CTF) approximation \cite{avargel2007,talmon2009} in the STFT domain (Block \ding{172} and \ding{173} in Fig.~\ref{figfc}). The CTF is a convolutive filter on the STFT coefficients of source signal rather than the conventional multiplicative filter, thus it is a more accurate representation of the STFT-domain binaural signals than multiplicative transfer function (MTF) approximation. }  In \cite{mine2016ITASL} the DP-RTF was estimated at each frequency by solving a multi-dimensional linear equation built from the statistics of the binaural signals. Estimated DP-RTFs were then fed to the single-source localization method of \cite{deleforge2014mapping}. It was observed that the DP-RTF features outperform features based on MTF \cite{avargel2007spl}.
In \cite{mine2016ITASL} it was assumed that only one single source is active and, therefore, a unique linear equation is constructed at each frequency using all available time frames. 
\addnote[trivial]{1}{However, for multiple sources, successive time frames at a given frequency may not belong anymore to a single source, and one has to enforce the WDO assumption. At each frequency, the multi-dimensional linear equation used for estimating the DP-RTF is now constructed from a frame region (a set of continuous frames) where only one source is assumed to be active. 
The extension of WDO assumption to frame regions is far from being trivial due to the overlap of multisources. A consistency-test algorithm is thus proposed to verify whether a frame region is associated with a single source or not  (Block \ding{174}  in Fig.~\ref{figfc}). } If so, a \textit{local} DP-RTF estimation is obtained by solving this local equation, otherwise this frame region is discarded.
Applying this principle to many different regions over the entire binaural power spectrogram leads to a set of DP-RTF estimates, each one assumed to correspond to one of the sources. In practice, sets of continuous frames associated with a single source at a given frequency widely exist due to the speech sparsity in the STFT representation. These estimated DP-RTFs are suitable for the proposed CGMM clustering framework: Predicted DP-RTFs (which are the means of the CGMM) are calculated offline from a reverberation-free propagation model, for instance head-related transfer functions (HRTFs) since we use recordings from either a dummy-head or a robot head (Block \ding{175}  in Fig.~\ref{figfc}). Then, the measured and predicted DP-RTF features are provided to the CGMM penalized likelihood maximization procedure. This procedure outputs the optimized CGMM component weights for all predefined candidate positions, from which source localization is finally performed using a peak selection routine (Block \ding{178}  in Fig.~\ref{figfc}). Overall, the proposed method leads to an efficient multiple-source localization method in the presence of noise and reverberations.

The remainder of the paper is organized as follows. The multiple-source localization method based on CGMM with maximization of penalized likelihood is described in Section~\ref{sec:localization}.
The estimation of DP-RTF from the microphone signals for the case of multiple speakers is presented in Section~\ref{sec:dprtf}.  
Experiments with both simulated and real data are presented in Section~\ref{sec:experiments}. Section~\ref{sec:conclusion} concludes the work.

\section{Multiple Sound Source Localization}\label{sec:localization}

\subsection{Speech Mixtures and Binaural Features}
\label{subsec:bf}

We consider non-stationary source signals $s^i(n)$, e.g. speech, where $i\in[1,I]$ denotes the source index. The received binaural signals are 
\begin{equation}\label{eq:signal}
\begin{array}{l}
\tilde{x}(n) = x(n)+u(n) = \sum\nolimits_{i=1}^{I}a^i(n)\star s^i(n)+u(n), \\
\tilde{y}(n) = y(n)+v(n) = \sum\nolimits_{i=1}^{I}b^i(n)\star s^i(n)+v(n),
\end{array}
\end{equation}
where $x(n)$ and $y(n)$ are the speech mixtures, $u(n)$ and $v(n)$ are the microphone noise signals, $a^i(n)$ and $b^i(n)$ are the binaural room impulse responses (BRIR) from source to microphone, and $\star$ denotes convolution. 
The binaural signals are transformed into the time-frequency (TF) domain by applying the STFT. As mentioned above, many types of binaural features can be extracted in the TF domain. 
Let $c_{p,k}$ denote the complex-valued binaural features of interest, where $p\in[1,P]$ is the frame index, and $k\in[0,K-1]$ is the frequency index. The nature of $c_{p,k}$, namely \addnote[eq22]{1}{DP-RTF features and their estimation from binaural signals are presented in Section~\ref{sec:dprtf}, more specifically they are computed in~(\ref{eq:cpk})}.
Based on the WDO assumption, a $c_{p,k}$ feature is associated with a single source. However, in practice, some of the TF bins are dominated by noise or by the presence of several sources, and hence they should not be considered by the clustering process.
Let $\mathcal{P}_k$ denote the set of frame indexes, at frequency $k$, that are associated with a single source. 
Let $\mathcal{C}=\{\{c_{p,k}\}_{p\in\mathcal{P}_k}\}_{k=0}^{K-1}$ denote the set of features over all frequencies and available frames. The procedure of selecting ``reliable'' features (i.e. generating $\mathcal{C}$) will also be detailed in Section~\ref{sec:dprtf}.
In this section, we exploit $\mathcal{C}$ to perform source localization. 
The multi-source localization problem is first cast into a probabilistic clustering problem using a complex-Gaussian mixture model (CGMM).

\subsection{Clustering-Based Localization}

In order to group $c_{p,k}$ features into several clusters and hence to achieve multiple-source localization, we adopt the complex-Gaussian mixture model (CGMM) formulation proposed in \cite{dorfan2015}. 
Each CGMM component corresponds to a candidate source position on a predefined grid. Source counting and localization are based on the selection of those components having the highest weights. 
In \cite{dorfan2015} several pairs of microphones are used so that two-dimensional (2D) localization on a 2D regular grid can be achieved. In this paper, we focus on using a single microphone pair and thus we can only estimate the sources' azimuths \cite{mandel2010,raspaud2010,may2011,woodruff2012}. The extension to several microphone pairs is straightforward.
We define a set $\mathcal{S}$ of $S$ candidate azimuths regularly placed on a circular grid. In the remainder, $s\in\mathcal{S}$ denotes a candidate azimuth.\footnote{For convenience $s$ can indifferently denote a source azimuth or an index of this azimuth within the grid, arbitrarily set from 1 to $S$.} 
The probability of an observed binaural feature $c_{p,k}\in\mathbb{C}$, given that it is emitted by a sound source located at $s$, is assumed to be drawn from a complex-Gaussian distribution with mean $c_k^{s}\in\mathbb{C}$ and variance $\sigma^2\in\mathbb{R}$:
\begin{align}
\label{eq:obs-like}
 P(c_{p,k} | s) = \mathcal{N}_c(c_{p,k};c_k^{s},\sigma^2) = \frac{1}{\pi\sigma^2} \exp\left(-\frac{|c_{p,k}-c_k^{s}|^2}{\sigma^2}\right).
\end{align}
The mean $c_k^{s}$ is the predicted binaural feature at frequency $k$ as provided by a direct-path propagation model. 
The latter can be derived from the geometric relationship between the microphones and the source candidate position. 
If an acoustic dummy head is used for the binaural recordings, as will be the case in our experiments, the head-related transfer function (HRTF) of the dummy head is used to predict the means $c_k^{s}$ by taking the HRTF ratio between channels,
for each grid point $s$ and for each frequency $k$.

We now consider the grid of all possible locations, in which case the probability of a binaural feature, given the grid locations, is drawn from a CGMM:
\begin{equation}
\label{eq:CGMM}
P (c_{p,k}| \mathcal{S}) = \sum_{s=1}^S\alpha_{s}\mathcal{N}_c(c_{p,k};c_k^{s},\sigma^2),
\end{equation}
where $\alpha_s \geq 0$ is the prior probability that the binaural feature is drawn from the $s$-th component, namely the prior probability that the source is located at $s$, with $\sum_{s=1}^S\alpha_{s}=1$. In the present work, $\alpha_s$ is referred to as the component weight. Let us denote the vector of weights with $\alphavect=[\alpha_1, ..., \alpha_S]\tp$.
Since the mixture means are determined based on the source-sensor geometry, and the variance is set to an empirical value $\sigma^2$ common to all components and all frequencies,\footnote{This was reported as a relevant choice in \cite{dorfan2015}, and our experiments confirmed that a constant variance outperforms other mechanisms, such as setting the variance to be candidate-dependent (i.e. $\sigma_s^2$), or frequency-dependent ($\sigma_k^2$), or both ($\sigma_{k,s}^2$).} the components of $\alphavect$ are the only free model parameters. 

Assuming that the observations in $\mathcal{C}$ are independent, the corresponding log-likelihood function (as a function of $\alphavect$) is given by:
\begin{align}\label{eq:loglik}
 \log \mathcal{L}(\mathcal{C} | \alphavect) = \sum_{k=0}^{K-1}\sum_{p\in\mathcal{P}_k}\text{log}\Big(\sum_{s=1}^S\alpha_{s}\mathcal{N}_c(c_{p,k};c_k^{s},\sigma^2)\Big).
\end{align}
Multiple-source localization amounts to the maximization of the log-likelihood \eqref{eq:loglik}.
Importantly, the model above integrates the binaural features of all frequencies by sharing the weights over frequencies, and considers as many components as grid points. 
Intuitively, after maximization of \eqref{eq:loglik}, an active speaker location corresponds to a component with a large weight. 
In practice, a plot of the weights as a function of azimuth indeed exhibits a quite smooth curve with a few peaks that should correspond to active speakers, see Section~\ref{sec:experiments}. 
Therefore, the detection and localization of active speakers could be jointly carried out by selecting the components with the largest weights. 
A simple strategy would consist of selecting the peaks that are above a threshold, as done in \cite{dorfan2015}, or of selecting the $N_s$ largest peaks if the number of active sources $N_s$ is known in advance. However, spurious peaks often appear, due to, e.g., reverberated phantom sources, corrupting the source detection and localization. 
In the next subsection we propose a penalized maximum likelihood estimator, to enforce a sparse solution for $\alphavect$ and remove such spurious peaks. 

\subsection{Penalized Maximum Likelihood Estimation}
\label{subsection:penalized-LME}

Let $C=|\mathcal{C}|$ denote the cardinality of $\mathcal{C}$, namely the number of binaural observations. We note that \eqref{eq:loglik} can be written as:
\begin{equation}
\label{eq:likelihood-rewritten}
 \log \mathcal{L}(\mathcal{C} | \alphavect) = \sum_{c=1}^C \log \Big( \sum_{s=1}^S g_{cs} \alpha_s \Big) = \mathbf{1}_C\tp \log ( \Gmat \alphavect ),
\end{equation}
where $\mathbf{1}_C$ denotes a vector in $\mathbb{R}^C$ with all entries set to 1, $\Gmat\in\mathbb{R}^{C\times S}$ is the matrix of probabilities \eqref{eq:obs-like} reorganized so that each row $\gvect_c$ of $\Gmat$ corresponds to an observation in $\mathcal{C}$ and each column corresponds to a candidate source position, and where we used the notation:
\[
\log (\Gmat \alphavect) = \big[ \log (\gvect_1 \alphavect), \dots, \log (\gvect_c \alphavect), \dots, \log (\gvect_C \alphavect) \big]\tp.
\]
Then, the maximization of the log-likelihood \eqref{eq:loglik} can be written as the following convex optimization problem:
\begin{align}\label{eq:cov}
\text{minimize} \quad  & -\mathbf{1}_C\tp \text{log}(\mathbf{G}\alphavect) \nonumber \\ 
 \text{s.t.} \quad & -\alphavect \preceq \mathbf{0}_S, \quad \mathbf{1}_S\tp \alphavect =1,    
\end{align}
where $\mathbf{0}_S$ denotes a vector in $\mathbb{R}^S$ with all entries set to zero, and $\preceq$ denotes entry-wise vector inequality.
This convex optimization problem with equality and inequality constraints can be solved by the primal-dual interior-point method (PDIPM) \cite{boyd2004}, which will be described in Section~\ref{pdip}.
\addnote[covem]{1}{This optimization problem has the same solution as the original problem of maximizing the log-likelihood (4). 
However, in the following, we introduce a regularization term to impose the sparsity of $\alphavect$, which can be can be easily added to (6), but cannot be easily added to (4) within an EM algorithm.}

We remind that the parameter $\alpha_s$ is the prior probability of having an active source at location $s$. In practice, the number of active speakers is much lower than the number of candidate locations on the grid. One may consider a grid with tens or hundreds of source locations, but only a handful of this locations correspond to actual sources. Therefore, we may seek a sparse vector $\alphavect$ i.e. with only a few nonzero entries.
To enforce the sparsity of $\alphavect$ we propose to add a penalty term to the objective function in (\ref{eq:cov}). The entries of $\alphavect$ are probability masses of a discrete random variable. 
Generally, the sparser the vector, the smaller entropy $H(\alphavect)=-\alphavect\tp \text{log}(\alphavect)$ is. Therefore, the entropy may be used as the required penalty. A sparse solution for $\alphavect$ can be obtained by solving the following optimization problem:
\begin{align}\label{eq:ccp}
 \text{minimize} \quad &-\frac{1}{C}\mathbf{1}_C\tp \text{log}(\mathbf{G}\alphavect)-\gamma\alphavect\tp \text{log}(\alphavect) \nonumber \\ 
 \text{s.t.} \quad & -\alphavect \preceq \mathbf{0}_S, \quad \mathbf{1}_S\tp \alphavect =1   
\end{align}
where $\frac{1}{C}$ plays the role of a normalization factor, and $\gamma$ is an empirical parameter that enables to control the trade-off between the log-likelihood and the entropy. 

The entropy $-\alphavect\tp \text{log}(\alphavect)$ is a concave function. 
Thence the problem can be solved via a convex-concave procedure (CCP) \cite{yuille2003}. 
 To solve the CCP, an iterative method is proposed in \cite{smola2005, lipp2016}. At each iteration, the concave function is approximated by its first-order Taylor expansion, so that the convex-concave function becomes a convex function.
The derivative of the entropy w.r.t. $\alphavect$ is  $-(1+\text{log}(\alphavect))$ and the first-order Taylor expansion at $\tilde{\alphavect}$ is
\begin{align}
 T_{H}(\alphavect,\tilde{\alphavect})=-\tilde{\alphavect}\tp \text{log}(\tilde{\alphavect})-(\alphavect-\tilde{\alphavect})\tp(1+\text{log}(\tilde{\alphavect})). \nonumber
\end{align}
The solution to (\ref{eq:ccp}) is summarized in Algorithm \ref{alg:CCP} which is referred to as EP-MLE (entropy-penalized maximum likelihood estimator). A convergence proof of this procedure is provided in \cite{smola2005, lipp2016}. 
Subproblem~(\ref{eq:subcov}) is a convex optimization problem with equality and inequality constraints and, again, it is solved with PDIPM.
The algorithm is stopped when the decrease of the objective function (\ref{eq:ccp}) from one iteration to the next is lower than a threshold $\delta$.  
CCP can have (many) local minima, therefore the initialization is important for searching the global minimum, just as for EM algorithms. If $\gamma$ is small, we assume that the global minimum is in the close proximity of the minimum of (\ref{eq:cov}).
Therefore, the initialization of Algorithm~\ref{alg:CCP} is set as the solution of (\ref{eq:cov}), obtained with PDIPM. 

\begin{algorithm}
\caption{\label{alg:CCP} Concave-convex minimization}
\begin{algorithmic} 
 \STATE Set $m=0$, initialize $\alphavect^{(0)}$ with the solution of (\ref{eq:cov}).
 \REPEAT 
 \STATE 1 Set $m := m+1$
 \STATE 2 Solve the convex optimization problem:
 \begin{align}\label{eq:subcov}
   & \alphavect_{opt} = \mathop{\textrm{argmin}}_{\alphavect} \{-\frac{1}{C}\mathbf{1}_C\tp \text{log}(\mathbf{G}\alphavect)+\gamma T_{H}(\alphavect,\alphavect^{(m-1)})\} \nonumber \\ 
 & \text{s.t.} \quad  -\alphavect \preceq \mathbf{0}_S, \quad \mathbf{1}_S\tp \alphavect =1   
 \end{align}  
 \STATE 3 Set $\alphavect^{(m)} := \alphavect_{opt}$
 \UNTIL Convergence
\end{algorithmic}
\end{algorithm}

\vspace{-0.2cm}
\subsection{The Primal-Dual Interior-Point Method}
\label{pdip}

We follow \cite{boyd2004} to solve for both (\ref{eq:cov}) and (\ref{eq:subcov}). \cite{boyd2004}
provides a general optimization algorithm for a convex objective function $f_0$ with a set of inequality constraints of the form $f \preceq \mathbf{0}$ and an affine equality constraint. 
Here $f_0(\alphavect)$ is the objective function in (\ref{eq:cov}) or (\ref{eq:subcov}), and $f(\alphavect)=-\alphavect$. 
\addnote[dgap]{1}{It is obvious that there exist feasible points for the convex problem (\ref{eq:cov}) and (\ref{eq:subcov}), namely the Slater's constraint qualification is satisfied. 
Therefore, the \emph{strong duality} holds for the present problems, in other words, the optimal duality gap is 0.}

\addnote[logbarrier]{1}{PDIPM makes the inequality constraints implicit in the objective function by applying the logarithmic barrier function. As for an inequality constraint $f\le0$, the logarithmic barrier  
\begin{align}
 \hat{I}_{\_}(f) = -(1/t)\text{log}(-f) \nonumber
\end{align}
is added to the objective. $\hat{I}_{\_}(f)$ takes the value $\infty$ for $f>0$ to penalize the objective. The logarithmic barrier is desirable due to its convexity and differentiability.  
Here $t$ sets the accuracy of the logarithmic barrier approximation, the larger $t$, the better the approximation.}

The optimization can be expressed as solving the Karush-Kuhn-Tucker (KKT) conditions:
\begin{align}\label{eq:kkt}
 r_t(\alphavect,\lambdavect,\nu)=
 \begin{bmatrix}
\nabla f_0(\alphavect)-\lambdavect+\nu\mathbf{1}_S \\
\text{diag}(\lambdavect)\alphavect-(1/t)\mathbf{1}_S\\
\mathbf{1}_S\tp \alphavect-1 \\
\end{bmatrix}
=\mathbf{0}
\end{align}
where $\lambdavect \in \mathbb{R}^S$ and $\nu \in \mathbb{R}$ are auxiliary variables that originate in the use of the Lagrange multiplier associated with the inequality and equality constraints, respectively. 
The nonlinear KKT conditions can be solved by Algorithm~\ref{alg:PDIPM}, with the update rule in Step~4 \addnote[newton]{1}{given by the Newton method}:
\begin{align}\label{eq:upd}
\begin{bmatrix}
\alphavect^{(n+1)} \\
\lambdavect^{(n+1)} \\
\nu^{(n+1)} \\
\end{bmatrix}
= &
\begin{bmatrix}
\alphavect^{(n)} \\
\lambdavect^{(n)} \\
\nu^{(n)} \\
\end{bmatrix}
-
 \begin{bmatrix}
\nabla^2f_0(\alphavect^{(n)}) & -\mathbf{I} & \mathbf{1}_S \\
\text{diag}(\lambdavect^{(n)}) & \text{diag}(\alphavect^{(n)}) & \mathbf{0}_S \\
\mathbf{1}_S\tp & \mathbf{0}_S\tp & 0 \\
\end{bmatrix}^{-1} \nonumber \\
\times &
\begin{bmatrix}
\nabla f_0(\alphavect^{(n)})-\lambdavect^{(n)}+\nu\mathbf{1}_S \\
\text{diag}(\lambdavect^{(n)})\alphavect^{(n)}-(1/t^{(n)})\mathbf{1}_S\\
\mathbf{1}_S\tp \alphavect^{(n)}-1 \\
\end{bmatrix} 
\times \zeta^{(n)} 
\end{align}    
where $^{(n)}$ denotes the iteration index, $\mathbf{I}$ is the identity matrix, and
$\zeta^{(n)}$ is the step-length.  
In the present study, the $j$th entry of the derivative vector of $f_0(\alphavect)$ is given by:
\begin{equation}
 \nabla f_0(\alphavect)_j =
\begin{cases} -\sum_{i=1}^C\frac{g_{ij}}{\sum_{j=1}^Sg_{ij}\alpha_j}, & \hspace{-30mm} \text{for~(\ref{eq:cov})} \\
-\sum_{i=1}^C\frac{g_{ij}}{\sum_{j=1}^Sg_{ij}\alpha_j}- \\
\hspace{2mm}\gamma(1+\text{log}(\alpha_j^{(m-1)})), \ \text{for~(\ref{eq:subcov}) (at iteration $m$)}& 
\end{cases}
\end{equation}
where $g_{ij}$ is the $(i,j)$-th entry of $\mathbf{G}$. For both (\ref{eq:cov}) and (\ref{eq:subcov}), the $(j_1,j_2)$-th entry of the Hessian matrix is:
\begin{equation}
 \nabla^2f_0(\alphavect)_{j_1j_2}=\sum_{i=1}^C\frac{g_{ij_1}g_{ij_2}}{(\sum_{j=1}^Sg_{ij}\alpha_j)^2}.
\end{equation}
Note that the update rule (\ref{eq:upd}) integrates the fact that the derivative of the inequality function $f(\alphavect)$ is $\nabla f(\alphavect)=-\mathbf{I}$ and that the Hessian matrix of one inequality function $f_s(\alphavect)=-\alpha_s$ is $\nabla^2f_s(\alphavect)=\mathbf{0}$ for $s \in [1,S]$.

\addnote[pdipm1]{1}{In Algorithm~\ref{alg:PDIPM}, the primal variable and dual variables are simultaneously updated, and the so-called surrogate duality gap $\hat{\eta}^{(n)}$ is decreasing with the iterations. }
 Correspondingly, the parameter $t$ is increased by the factor $\mu$ (a positive value of the order of 10) with respect to $\hat{\eta}^{(n)}$. 
The line search method for setting the step-length $\zeta^{(n)}$ (Step 3) is briefly summarized in Algorithm~\ref{alg:LSM}. 
Basically, the step-length is set as the largest value that makes the updated variables satisfy the three conditions (i)~the dual variable $\lambdavect$ is nonnegative, (ii)~the inequality constraint is satisfied, and (iii)~the overall KKT residual is decreased. 
In this work, the backtracking parameters $\beta$ and $\eta$ of Algorithm~\ref{alg:LSM} are set to 0.5 and 0.05, respectively. 
In the convergence criterion of Algorithm~\ref{alg:PDIPM}, \addnote[pdipm2]{1}{the surrogate duality gap $\hat{\eta}^{(n)}$ is compared with a small threshold $\epsilon$ (close to the optimal duality gap, i.e., 0) to guarantee the optimization. } The two other criteria are set to guarantee the feasibility of the variables ($\epsilon_{feas}$ is also a small arbitrary threshold). For solving (\ref{eq:cov}), a good initialization is to set $\alphavect^{(0)}=(1/S)\mathbf{1}_S$, $\lambdavect^{(0)}$ to an arbitrary positive vector ($10\cdot\mathbf{1}_S$ in this paper), and $\nu^{(0)}$ to an arbitrary value (0 in this paper).
For solving (\ref{eq:subcov}) in Algorithm~\ref{alg:CCP}, the initialization is set as the solution of the previous iteration.  
Finally, as already mentioned, Algorithm~\ref{alg:CCP} is initialized by the solution of (\ref{eq:cov}).
 
 \begin{algorithm}
 \caption{\label{alg:PDIPM} Primal-dual interior-point}
\begin{algorithmic} 
 \STATE Set $n=0$, Initialize $-\alphavect^{(0)} \preceq \mathbf{0}$, $\lambdavect^{(0)}\succ\mathbf{0}$, $\nu^{(0)}$.
 \REPEAT 
 \STATE 1 Compute $\hat{\eta}^{(n)}=\{\alphavect^{(n)}\}\tp \lambdavect^{(n)}$,
 \STATE 2 Set $t^{(n)}:=\mu S/\hat{\eta}^{(n)}$,
 \STATE 3 Line search the step-length $\zeta^{(n)}$ (Algorithm 3),
 \STATE 4 Update variables with (\ref{eq:upd}).
 \UNTIL $\hat{\eta}^{(n)}\le\epsilon$, $\parallel \mathbf{1}_S\tp \alphavect^{(n)}-1 \parallel_2 \le \epsilon_{feas}$, and \\ \hspace{10mm} $\parallel \nabla f_0(\alphavect^{(n)})-\lambdavect^{(n)}+\nu\mathbf{1}_S\parallel_2 \le \epsilon_{feas}$ 
\end{algorithmic}
\end{algorithm}

 \begin{algorithm} \caption{\label{alg:LSM} Line search}
\begin{algorithmic} 
 \STATE  Compute $\zeta^{\text{max}}=\text{sup}\{\zeta^{(n)}\in[0,1]|\lambdavect^{(n+1)}\succeq \mathbf{0}_S\}$,
 i.e. the largest $\zeta$ value that makes the updated $\lambdavect$ value nonnegative.   
 \STATE  Set $\zeta^{(n)}:=0.99\zeta^{\text{max}}$.
 \REPEAT 
 \STATE Set $\zeta^{(n)}:=\beta\zeta^{(n)}$
 \UNTIL $-\alphavect^{(n+1)} \preceq \mathbf{0}_S$ (i.e. the inequality constraint holds) and  $\parallel  r_t(\alphavect,\lambdavect,\nu)^{(n+1)}\parallel_2 \le (1-\eta\zeta^{(n)})\parallel  r_t(\alphavect,\lambdavect,\nu)^{(n)}\parallel_2$ (i.e. the overall KKT residual is decreased).
 \end{algorithmic}
\end{algorithm}


\section{Direct-Path Estimation for Multiple Speakers}\label{sec:dprtf}

In this section we propose to estimate the direct-path relative transfer function (DP-RTF) for multiple speakers, which is an extension of the single-speaker case \cite{mine2016ITASL}. 
The rationale of using the DP-RTF is twofold. First, it is robust to noise and reverberations and, second, it is a well-suited binaural feature to be used within the complex-valued generative model \eqref{eq:CGMM}. For clarity, we first briefly present the single-speaker case \cite{mine2016ITASL}, and then we move to the multiple-speaker case.

\subsection{DP-RTF Estimation for a Single Speaker}
In the case of a single speaker, the noise-free received binaural signals are
\begin{align}\label{xn}
 x(n)=s(n)\star a(n), \quad y(n)=s(n)\star b(n).
\end{align}
In the STFT domain, the MTF approximation is only valid when the impulse responses $a(n)$ and $b(n)$ are short, relative to the STFT window. 
To represent a linear filter with long impulse response in the STFT domain more accurately, the cross-band filters were introduced \cite{avargel2007,gilloire1992}, 
and a CTF approximation is further introduced and used in \cite{talmon2009} to simplify the analysis.
Let $N$ and $L$ denote the size and the shift of the STFT window, respectively. Following the CTF, $x(n)$ is approximated in the STFT domain by:
\begin{align}
\label{eq:xpk3}
 x_{p,k} = \sum_{p'=0}^{Q-1} s_{p-p',k}a_{p',k}= s_{p,k}\star a_{p,k},  
 \end{align}
\addnote[conv]{1}{where $x_{p,k}$ and $s_{p,k}$ are the STFT of $x(n)$ and $s(n)$, respectively, $a_{p,k}$ is the CTF of the filter, and where the convolution $\star$ is executed with respect to the frame index $p$ }.
The number of CTF coefficients $Q$ is related to the reverberation time.
The first CTF coefficient $a_{0,k}$ can be interpreted as the $k$-th coefficient of the Fourier transform of the impulse response segment $a(n)|_{n=0}^{N-1}$. 
This holds whatever the actual size of $a(n)$, including if this size is much larger than the STFT window length $N$.
Without loss of generality, we assume that the room impulse response $a(n)$ begins with the impulse response of the direct-path propagation.
If the frame length $N$ is properly chosen, $a(n)|_{n=0}^{N-1}$ is thus composed of the direct-path impulse response and possibly of a few reflections. 
Hence we refer to $a_{0,k}$ as the direct-path acoustic transfer function (ATF). A similar statement holds for $b(n)$ and its corresponding direct-path ATF $b_{0,k}$.  By definition, the DP-RTF is given by $\frac{b_{0,k}}{a_{0,k}}$.
We remind that the direct-path propagation model in general, and the DP-RTF in particular, have proven to be relevant for sound-source localization.

Based on the cross-relation method \cite{xu1995}, using the CTF model of two channels in the noise-free case we have: $x_{p,k}\star b_{p,k}=y_{p,k}\star a_{p,k}$.
Dividing both sides by $a_{0,k}$ and reorganizing the terms in vector form we can write:
 \begin{align}\label{eq:zpk}
 y_{p,k} = \mathbf{z}_{p,k}\tp \; \mathbf{g}_k,
 \end{align}
where
\begin{align}
 \mathbf{z}_{p,k} &= [x_{p,k},\dots,x_{p-Q+1,k},y_{p-1,k},\dots,y_{p-Q+1,k}]\tp \nonumber \\
\mathbf{g}_k &=\left[\frac{b_{0,k}}{a_{0,k}},\dots,\frac{b_{Q-1,k}}{a_{0,k}},-\frac{a_{1,k}}{a_{0,k}},\dots,-\frac{a_{Q-1,k}}{a_{0,k}}\right]\tp. \nonumber
\end{align}
We see that the DP-RTF appears as the first entry of the reverberation model $\mathbf{g}_k$. 
By multiplying both sides of (\ref{eq:zpk}) with $ y_{p,k}^*$ (the complex conjugate of $ y_{p,k}$) and by taking the expectation (in practice averaging the corresponding power spectra over consecutive $D$ frames), 
we obtain:
 \begin{align}\label{hatphi}
 \hat{\phi}_{yy}(p,k) = \hat{\phivect}_{zy}\tp(p,k) \: \mathbf{g}_k,
 \end{align}
where $\hat{\phi}_{yy}(p,k)$ is the power spectral density (PSD) estimate of $y(n)$ at TF bin $(p,k)$, and $\hat{\phivect}_{zy}(p,k)$ is a vector composed of cross-PSD terms between the elements of $\mathbf{z}_{p,k}$ and $y_{p,k}$.
 

As for the noisy case, an inter-frame spectral subtraction algorithm can be used for noise suppression, e.g. \cite{mine2016ITASL}:
The auto- and cross-PSD of a frame with low speech power are subtracted from the auto- and cross-PSD of a frame with high speech power. 
Due to the stationarity of noise and the non-stationarity of speech, the resulting power spectra estimates, $\hat{\phi}_{yy}^s(p,k)$ and $\hat{\phivect}_{zy}^s(p,k)$, have low noise power and high speech power. 
Let $\mathcal{P}_k^s$ be the set of frame indices with high-speech power (at frequency $k$). 
After the spectral subtraction, we have:
\begin{align}\label{eq:gpk}
 \hat{\phi}_{yy}^s(p,k) = \hat{\phivect}_{zy}^s(p,k)\tp  \mathbf{g}_{k}+e(p,k), \quad p\in\mathcal{P}_k^s,
\end{align}
with $e(p,k)$ denoting the residual noise of the spectral subtraction procedure. 
Using the frames indexed in $\mathcal{P}_k^s$, a set of linear equations can be built and solved, yielding an estimate $\hat{\mathbf{g}}_k$ of $\mathbf{g}_k$ and its first component is the estimated DP-RTF.

\subsection{DP-RTF Estimation for Multiple Speakers}\label{sec:dprtf_est}
As just summarized, all the frames in $\mathcal{P}_k^s$ can be used to construct a DP-RTF estimate in the case of a single speaker. 
This is no more valid in the case of multiple speakers, since the frames in $\mathcal{P}_k^s$ do not necessarily correspond to the same source. Hence the DP-RTF estimation method must be reformulated in the case of multiple emitting sources. By applying the STFT to (\ref{eq:signal}), the recorded binaural signals write:
\begin{equation}\label{eq:xypk}
\begin{array}{l}
\tilde{x}_{p,k} = x_{p,k}+u_{p,k} = \sum_{i=1}^I s_{p,k}^i\star a_{p,k}^i+u_{p,k},  \\
\tilde{y}_{p,k} = y_{p,k}+v_{p,k} = \sum_{i=1}^I s_{p,k}^i\star b_{p,k}^i+v_{p,k}.
\end{array}
 \end{equation}
Without any additional assumption, (\ref{eq:gpk}) does not generalize to multiple sources, and thus we cannot directly estimate the DP-RTF associated to each source using the statistics of the mixture signals $x(n)$ and $y(n)$ measured on any arbitrary set of frames. To exploit the above results, we resort to the WDO assumption, i.e. we assume that in a small region of the TF plane only one source is active. 
Based on this assumption, the DP-RTF in a given TF bin is assumed to correspond to at most one active source. In the following, we thus choose to estimate the DP-RTF for each TF bin. We first formalize this estimate based on the above results. Then we discuss the assumptions for which this estimate is valid and we propose a consistency test to either select the DP-RTF in a given TF bin as a valid estimate for one of the sources or reject it (i.e. we consider that it is not a valid DP-RTF estimate of one of the sources). 
Using the WDO assumption, and defining:
\begin{align}
\mathbf{g}^i_k &=\left[\frac{b_{0,k}^i}{a_{0,k}^i},\dots,\frac{b_{Q-1,k}^i}{a_{0,k}^i},-\frac{a_{1,k}^i}{a_{0,k}^i},\dots,-\frac{a_{Q-1,k}^i}{a_{0,k}^i}\right]\tp, i \in [1,I], \nonumber
\end{align}
whose first entry is the DP-RTF of source $i$, we have a possible value $\mathbf{g}_{p,k} \in\{\mathbf{g}_k^i\}_{i=1}^I$ at each STFT bin.
In order to estimate $\mathbf{g}_{p,k}$, an equation of the form (\ref{eq:gpk}) has to be constructed for a set of frames corresponding to a single source. Let us consider such a set of $O$ consecutive frames to form:
 \begin{align}\label{eq:Phin}
 \hat{\phivect}_{yy}^s(p,k) = \hat{\Phimat}_{zy}^s(p,k)\mathbf{g}_{p,k}+\mathbf{e}(p,k), \ \mathbf{g}_{p,k}\in\{\mathbf{g}_k^i\}_{i=1}^I
 \end{align}
 where 
\begin{align}
 \hat{\phivect}_{yy}^s(p,k)&=[\hat{\phi}_{yy}^{s}(p-O+1,k),\dots,\hat{\phi}_{yy}^{s}(p,k)]\tp, \nonumber \\
 \hat{\Phimat}_{zy}^s(p,k)&=[\hat{\phivect}_{zy}^s(p-O+1,k),\dots,\hat{\phivect}_{zy}^s(p,k)]\tp, \nonumber \\
 \mathbf{e}(p,k) &= [e(p-O+1,k),\dots,e(p,k)]\tp, \nonumber
\end{align}
are $O\times1$ vector, $O\times(2Q-1)$ matrix and $O\times1$ vector, respectively.  
Note that (most of) the frames involved in the construction of $\hat{\phivect}_{yy}^s(p,k)$ and  $\hat{\Phimat}_{zy}^s(p,k)$ should have high-speech power, i.e. $[p-O, p]\subseteq \mathcal{P}_k^s$. Assume that $\mathbf{e}(p,k)$ is stationary and independent along frames.
Then if the matrix $\hat{\Phimat}_{zy}^s(p,k)$ is not underdetermined, i.e. $O\ge2Q-1$, an optimal estimation of $\mathbf{g}_{p,k}$ is given by the least square solution of (\ref{eq:Phin}):
\begin{align}\label{eq:ls}
 \hat{\mathbf{g}}_{p,k} = (\hat{\Phimat}_{zy}^s(p,k)^H\hat{\Phimat}_{zy}^s(p,k))^{-1}\hat{\Phimat}_{zy}^s(p,k)^H\hat{\phivect}_{yy}^s(p,k).
\end{align}
Let $\sigma_k^2$ denote the variance of the residual noise $e(p,k)$.
The covariance matrix of $\hat{\mathbf{g}}_{p,k}$ is $\sigma_k^2(\hat{\Phimat}_{zy}^s(p,k)^H\hat{\Phimat}_{zy}^s(p,k))^{-1}$ \cite{manolakis2005}, 
which obviously can be reduced by enlarging the number of equations, i.e. $O$.

To estimate the cross-PSD between $y_{p-Q+1,k}$ (or $x_{p-Q+1,k}$) and $y_{p,k}$, the past $D-1$ frames before the $(p-Q+1)$-th frame are employed.
Therefore, the STFT coefficients in the frame range $[p-Q-D+2,\ p]$ should be associated with a single active speaker. When considering $O$ consecutive frames, the past $Q+D-2$ frames before the $(p-O+1)$-th frame are employed to construct the earliest cross-PSD vector in (\ref{eq:Phin}), i.e. $\hat{\phivect}_{zy}^s(p-O+1,k)$. Therefore, for a correct estimation of the DP-RTF at TF bin $(p,k)$, the STFT coefficients at frequency $k$ in the frame range $[p-O-Q-D+3,\ p]$ should be associated with a single active speaker. 
In contrast, if the $O+Q+D-2$ consecutive speech frames used in the estimation of a DP-RTF at TF bin $(p,k)$ are composed of coefficients involving multiple active speakers, (\ref{eq:ls}) will not deliver a valid estimate of the DP-RTF, i.e. a DP-RTF estimate that corresponds to one and only one of the sources.
In other words, the present work requires a stricter WDO assumption than the original one \cite{rickard2002,yilmaz2004}, since at a given frequency bin $k$, we seek multiple continuous frames associated to a same single source.  

\begin{figure}[t]
\centering
\subfloat[Magnitude of STFT coefficient]{\includegraphics[width=1\columnwidth]{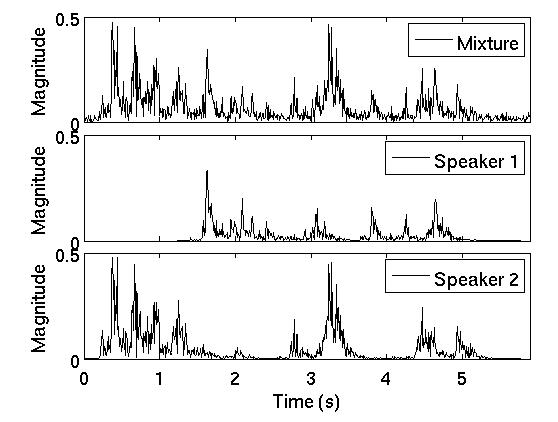}\label{2a}}  \\ 
\subfloat[Phase of DP-RTF estimate and consistency test]{\includegraphics[width=1\columnwidth]{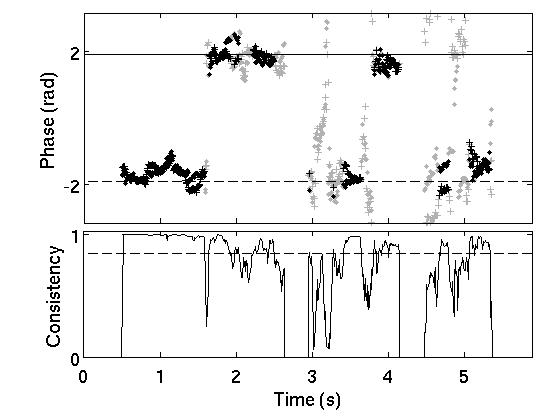}\label{2b}} 
\caption{\small{An example of multispeaker DP-RTF estimate at a given frequency ($2$~kHz). Binaural simulations with two speakers at $-40^\circ$ and $40^\circ$, SNR = $20$~dB, reverberation time $=$ 0.6~s (see the detailed dataset description in Section~\ref{exp:simdata}). Speaker 1 is active from $\approx\!1.5$~s to $\approx\!5$~s, while Speaker 2 is active all the time.
(a): Magnitude of STFT coefficient vs. time.  (b): Phase of DP-RTF estimates (top) and consistency test (bottom), both on the mixture signal. In the top figure, the dots represent the phase of the DP-RTF estimate, i.e. $\textrm{arg}[\hat{c}_{p,k}]$, and the cross points (+) represent the estimates after exchanging channels, i.e. $\textrm{arg}[1/\hat{c}'_{p,k}]$. 
The markers in black and grey indicate that the TF bins pass the consistency test or not, respectively. The solid line and dashed line denote the predicted phase computed from the HRTFs of Speaker 1 and Speaker 2, respectively.
In the bottom figure, the solid curve represents the similarity measure in (\ref{eq:similarity}), the dashed line is the threshold set to 0.85. 
Note that a zero similarity means that there is not enough frames with high speech power in the corresponding region to construct (\ref{eq:Phin}).}} 
\label{fig:feature}
\vspace{-0.7mm}
\end{figure}

\addnote[frame_region]{1}{In a scenario with multiple and simultaneous speech sources, the natural sparsity and the harmonic nature of speech spectra in the STFT domain make it common that at a given frequency a set of consecutive speech frames is dominated by a single active speaker. 
However, the amount of speech regions dominated by a single speaker is decreasing with an increasing number of sources $I$ and an increasing CTF length $Q$.
Fig.~\ref{2a} shows an example of two-speaker mixture at one given frequency (for instance $2$~kHz). 
It can be seen that the magnitude spectrum (at the selected frequency) of individual speech signals exhibits regions with large energy over numerous consecutive frames. This is expected to correspond to a signal harmonic. 
We observe that most regions are dominated by a single speaker, and as a result, the trajectory of the mixture magnitude coefficient ressembles the sum of the magnitude of the two individual speech signals. 
This indicates that the WDO assumption can be relaxed to a few hundreds of milliseconds.
Note that this mixture is just an example at one given frequency. The overlap between different sources could be much more (or less) than this illustrated example. 
Taking all the frequency bands into account, there exist a notable number of speech regions dominated by a single speaker for cases where i) the reverberation time is not very long, e.g. not longer than 0.7~s and ii) the number of sources is not very large, e.g. not larger than three.\footnote{The requirement of $O$ consecutive frames is to guarantee the least square problem~(\ref{eq:Phin}) to be not underdetermined. Based on the analysis of the cross-relation method in \cite{xu1995}, for the multichannel case, $O$ would be proportional to $\frac{1}{I-1}$. 
Therefore, $O$ could be reduced, namely WDO assumption could be relaxed, by increasing the number of channels. The case $I>2$ is beyond the scope of this work, and will be investigated in future work.}
For source localization, we only need a certain number of speech regions to be valid, rather than requiring most of the TF-bins to be valid, as is the case for binary-mask source separation. } In the next subsection, we propose a consistency test method to efficiently pick out the valid speech regions.

\subsection{Consistency Test}

A consistency test is proposed to check whether a continuous set of  $(O+Q+D-2)$ STFT coefficients at a given frequency $k$  are associated with a single active speaker or not. 
The principle is based on exchanging the roles of the two channels, since the DP-RTF between $b(n)$ and $a(n)$ is the inverse of the DP-RTF between $a(n)$ and $b(n)$. 
We thus define $\mathbf{g}'_{p,k}$ as the reverberation model that exchanges the roles of $a_{p,k}$ and $b_{p,k}$ in $\mathbf{g}_{p,k}$. 
If the STFT coefficients used to estimate $\hat{\mathbf{g}}_{p,k}$ and $\hat{\mathbf{g}}'_{p,k}$ are associated with a single speaker, (\ref{eq:zpk}) holds and the two corresponding DP-RTF estimates should be consistent. 
\addnote[ctcomment]{1}{Conversely, if the STFT coefficients are associated with more than one speaker, or only with reverberations, the estimations $\mathbf{g}_{p,k}$ and $\mathbf{g}'_{p,k}$ are both biased, with inconsistent bias values.
As a result, we should observe a discrepancy between the two estimated DP-RTF values.} 

In practice, let us denote by $\hat{c}_{p,k}$ and $\hat{c}'_{p,k}\in\mathbb{C}$ the first entry of $\hat{\mathbf{g}}_{p,k}$ and of $\hat{\mathbf{g}}'_{p,k}$ respectively, i.e. the DP-RTF estimates $\frac{b_{0,k}}{a_{0,k}}$ and $\frac{a_{0,k}}{b_{0,k}}$. 
We test the consistency by measuring the difference between $\hat{c}_{p,k}$ and $1/\hat{c}'_{p,k}$. To achieve a normalized difference measurement that allows us to easily set a reasonable test threshold, 
we define the vectors $\mathbf{c}_{1,p,k}=[1, \hat{c}_{p,k}]\tp$ and $\mathbf{c}_{2,p,k}=[1, 1/\hat{c}'_{p,k}]\tp$, where the first entry 1 can be interpreted as the DP-RTF corresponding to $\frac{a_{0,k}}{a_{0,k}}$.
The similarity, i.e. the cosine of the angle, of the two vectors:
\begin{align}\label{eq:similarity}
d_{p,k}= \frac{|\mathbf{c}_{1,p,k}\tp\mathbf{c}_{2,p,k}|}{\sqrt{\mathbf{c}_{1,p,k}\tp \mathbf{c}_{1,p,k}\mathbf{c}_{2,p,k} \tp \mathbf{c}_{2,p,k} }}
\end{align}
is a value in $[0,1]$, which is a good difference measurement. The larger $d_{p,k}$, the more consistant the reverberation model is. The consistency decision is made by comparing $d_{p,k}$ with a threshold $d_T$ (e.g. set to $0.85$).  

\addnote[ct]{1}{An example of consistency test is shown in Fig.~\ref{2b}. The test is applied to the mixture signal in Fig.~\ref{2a}. It can be seen that the phase of $\hat{c}_{p,k}$ and $1/\hat{c}'_{p,k}$ are close to each other, and are close to the predicted phase, for the frames dominated by a single speaker, e.g. within 0.3--1.5 s for Speaker 1. Correspondingly, the consistency measures are large (close to 1). 
For the regions that involve the two speakers, e.g. around 3.2 s, and the regions that mainly involve the reverberations, e.g. around 3.7 s, the two phase measures are far from the predicted phase, and are far from each other, thus the consistency measures are low. 
Eventually, the DP-RTF estimates that pass the consistency test are correctly selected, as shown by the black markers, which are close to the predicted value.}

Let $\mathcal{P}_k$ denote the set of frames indices that pass the consistency test for frequency $k$. 
Every DP-RTF estimation in $\mathcal{P}_k$ is first recalculated as $(\hat{c}_{p,k}+1/\hat{c}'_{p,k})/2$ to improve the estimate robustness. Finally it is normalized as 
\begin{align}\label{eq:cpk}
c_{p,k} = \frac{(\hat{c}_{p,k}+1/\hat{c}'_{p,k})/2}{1+|(\hat{c}_{p,k}+1/\hat{c}'_{p,k})/2|},
\end{align}
which is a complex number whose module is in the interval $[0,1]$. Each $c_{p,k}$ is assumed to be associated with a single speaker. We thus now have a set of normalized DP-RTF observations that are ready for clustering among sources.

\section{Experiments} \label{sec:experiments}

In this section, we present a series of experiments with simulated data and real data collected from a robotic head. 
We start by describing the experimental setup, and then give the experimental results and discussions.


\subsection{Experimental Setup}

\subsubsection{Blind and Semi-Blind Configurations}
Two configurations were tested, blind and semi-blind. In the blind configuration, the number of active sources $I$ and their locations are simultaneously estimated. 
\addnote[blind]{1}{Note that the term `blind' mainly refers to the unknown number of sources, and does not mean a complete blind configuration, for instance the HRTFs and reverberation time are known. }
Localization is conducted by selecting the local maxima in the set of CGMM weights that are above a threshold $\alpha_T$, i.e. we detect $\{\alpha | \alpha > \alpha_T, \alpha \in [\alpha_1, \dots, \alpha_S]\}$. In the semi-blind configuration, $I$ is assumed to be known and the source locations are detected by selecting the $I$ largest local maxima over the weights $[\alpha_1, \dots, \alpha_S]$.
The source location estimates are associated to the ground-truth source locations by looking for the correspondence that provides the overall lower mean absolute localization error (MAE) averaged across sources.
\addnote[peak]{1}{In general, blind localization is more difficult than semi-blind localization in terms of peak selection.} 


\subsubsection{Performance Metrics}
For both configurations, a source is then considered to be successfully localized if the difference between its actual azimuth and the estimated azimuth is not larger than a predefined threshold, empirically set to $15^{\circ}$.  Then, a new MAE is calculated for the successfully localized sources, which is the MAE in the results reported below. To further characterize the unsuccessful localizations in the blind configuration scenario, we also calculated: 
(i)~the missed detection (MD) rate defined as the percentage of sources that are present but not detected out of the total number of present sources; and (ii)~the false alarm (FA) rate defined as the percentage of sources that are detected although they are not actually present in the scene, out of the total number of sources. In the semi-blind configuration, we calculated the outlier rate, defined as the percentage of sources for which the azimuth error is larger than $15^{\circ}$ out of the total number of present sources (in short, the percentage of unsuccessfully localized sources). 
Note that, on one hand, an outlier indicates a missed detection of the corresponding true source, on the other hand, the outlier estimate itself is a false alarm. 

\subsubsection{Parameters Setting}

The signal sampling rate is 16 kHz. Only the frequency band from 0 to 4 kHz is considered for speech source localization, since this band concentrates the largest part of speech signals energy. The STFT frame length is set to $N=16$~ms (256~samples) with frame shift $L=8$~ms (128~samples). 
The CTF length $Q$ is set to correspond to $T_{60}/6$. The number of frames for the PSD estimate is $D=15$ (120~ms).
We set $O=3.5Q$ as a trade-off for ensuring a small variance of $\hat{\mathbf{g}}_{p,k}$, and the sparsity of the speech spectrum (one single active source) on a reasonable number of successive frames. 
The threshold for the consistency test is set to $d_T=0.85$.
The penalty factor $\gamma$ in (\ref{eq:ccp}) is set to $0.2$ as a good experimental trade-off between the log-likelihood and the entropy. The positive factor $\mu$ in Algorithm~\ref{alg:PDIPM} is set to $20$. 
The thresholds for the convergence criterion in Algorithms~\ref{alg:CCP} and \ref{alg:PDIPM} are set to $\delta=10^{-3}$ and $\epsilon=\epsilon_{feas}=10^{-6}$.
In the blind localization configuration, the threshold $\alpha_T$ for the local maximum selection corresponding to source detection is set to $0.05$, since this value was shown to provide a good trade-off between MD and FA.

\subsubsection{Simulated Binaural Data}\label{exp:simdata}
A set of BRIRs were generated with the ROOMSIM simulator \cite{campbell2004} combined with the head-related impulse responses (HRIRs) of the KEMAR dummy head \cite{gardner1995}. 
The simulated room is of dimension $5$~m $\times$ $8$~m $\times$ $3$~m. The dummy head is located at ($1$~m, $4$~m, $1.5$~m). 
Sound sources were placed in front of the dummy head with azimuths (relative to the dummy head center) varying from $-90^\circ$ to $90^\circ$, spaced by $5^\circ$ (hence $37$ azimuths), and an elevation of $0^\circ$. Five sets of $37$ binaural signals were generated by selecting $5$ different speech signals from the TIMIT dataset \cite{garofolo1988} and convolving each of these $5$ signals with each of the $37$ BRIRs.

We set the reverberation time to $T_{60} = 0.6$~s, which is quite notable. Accordingly, we set $Q=12$ ($96$ ms) and $O=42$.
Two dummy-head-to-source distances were simulated, namely $1$~m and $2$~m, for  which the direct-to-reverberant ratio (DRR) is about $0.5$~dB and $-5.5$~dB, respectively.
Localization of two and three speakers is considered. We generated $500$ mixtures for each case, by summing binaural signals randomly selected from the five groups, ensuring that the source directions are spaced by at least $15^\circ$. 
The noise signals were generated by mixing two types of noise with the same power: (i)~directional noise: white Gaussian noise emitted from the source point with azimuth of $120^\circ$, elevation of $30^\circ$ and distance-to-head of $2.2$~m, and (ii)~spatially uncorrelated white Gaussian noise. 
The composite noise signal was added to the speech mixture signals with signal-to-noise ratio (SNR) of either $30$~dB or $5$~dB.
The duration of each noisy speech mixture used for localization is of about $3$~s. 
Importantly, in these simulations, the predicted DP-RTF corresponding to the candidate source locations, i.e. the means of the CGMM components, are computed by using the anechoic HRIRs from \cite{gardner1995}, which ideally corresponds to the direct-path of the complete simulated propagation model (the BRIRs). The set of candidate locations $\mathcal{S}$ is composed of the $37$ azimuth values within $[-90^\circ, 90^\circ]$ taken every $5^\circ$. 

\subsubsection{Robotic Head Data}

\begin{figure}[t]
\centering
{\includegraphics[width=0.35\textwidth]{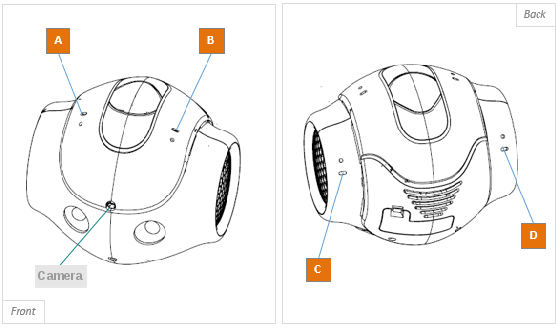}}
\caption{\small{The four-microphone robot head used in this paper.}} 
\label{fignao}
\vspace{-3mm}
\end{figure}

We also report  real-world experiments conducted using the head of the NAO humanoid robot (version 5), equipped with four nearly-coplanar microphones, see Fig.~\ref{fignao}. 
Elevation localization is here unreliable due to the coplanar microphone array. We used the two microphone pairs A-C and B-D to localize the azimuth relative to the NAO head.
The head has built-in fans nearby the microphones, hence the recorded data contain a notable amount of fan noise (aka ego-noise), which is stationary and spatially correlated \cite{loellmann2014}.

The data are recorded in an office room with $T_{60}=0.52$~s. Accordingly, we set here $Q=11$ ($88$ ms) and $O=38$.
The test dataset consists of long speech utterances  ($> 3$~s) from the TIMIT dataset, emitted by a loudspeaker. 
Two data sets are recorded with a robot-to-source distance of $1.5$~m and $2.5$~m, respectively (remember that DRR is related to the microphone-to-source distance). 
For each data set, $174$ speech utterances were emitted from directions uniformly distributed in the range $[-120^\circ,120^\circ]$ for azimuth, and $[-15^\circ, 25^\circ]$ for elevation.
The noise of recorded signals mainly corresponds to fan noise, the SNR is about $10$~dB. 
Two-speaker localization and three-speaker localization were considered. For each case, $200$ mixtures were generated by summing the sensor signals from two or three different directions. 
\addnote[noise]{1}{Note that this mixing  procedure sums the noise signals from each individual recording, which is different from what would be obtained with a real mixture recording. 
The summed noise has statistical property identical to the individual noises since latter are identically distributed and stationary, while the SNR is decreased. }    
The mixture signals were truncated to have a duration of $3$~s. 
The source azimuths are spaced by a random angle not lower than $15^\circ$.
The candidate azimuths $\mathcal{S}$ are here set to values within  $[-120^\circ, 120^\circ]$ with a $6^\circ$-step, hence there are $41$ candidate azimuths. 
As for the two microphone pairs, the predicted binaural features (CGMM mean) of the candidate azimuths were respectively computed by using the corresponding anechoic HRTFs. 
\addnote[naohrtf]{1}{The HRTFs and the predicted features are computed offline from HRIRs measured in laboratory: white Gaussian noise is emitted from a loudspeaker placed around NAO's head from each candidate direction, and the cross-correlation between the microphone and source signals yields the BRIR of each direction. In order to obtain anechoic HRIRs, the BRIRs are manually truncated before the first reflection. }

The information from the two microphone pairs was integrated into the localization model with the following procedure: 
1) binaural features are extracted independently from each of the two pairs, 
2) the Gaussian probabilities of the binaural features are computed using (\ref{eq:obs-like}) for each pair; note that the CGMM means $c_k^s$ are different for the two pairs, but the weights $\alphavect$ are of course the same,
3) we have an additional summation over the two pairs of features in the likelihood function~(\ref{eq:loglik}); this corresponds to have the Gaussian probabilities of the two pairs concatenated into a common matrix $\mathbf{G}$, and finally 4) execute the optimization procedure.

\subsection{Baseline Methods}

The results of the proposed method are compared with the results obtained with the four following baseline methods.

\subsubsection{Basic-CGMM} To test the relevance and efficiency of the entropy penalty, the results obtained with the same CGMM model, but solving the basic optimization problem (\ref{eq:cov}), i.e. without the entropy penalty, are compared. The same proposed DP-RTF feature is used here, and the peak counting threshold of the blind configuration is empirically set to $0.15$ to adjust the trade-off between MD and FA. 

\subsubsection{RTF-CT-CGMM} To test the efficiency of the proposed DP-RTF feature, the binaural RTFs with normalized amplitude of \cite{mohan2008} are tested for comparison. Here, a coherence test is used to search the TF bins which are supposed to be dominated by one active source. Note that the direct-path source and its reflections are considered as different sources, thence, the TF bins that pass the coherence test are supposed to be dominated by the direct-path signal of one active source.
The TF bins that have a coherence larger than a threshold (here set to $0.9$) are selected to provide RTF features. 
The proposed CGMM localization model is used. 
For the blind configuration, the peak counting threshold is set to $0.15$ as a good trade-off between MD and FA.
Note that, only the TF bins that have a high speech power are considered for the coherence test. The inter-frame spectral subtraction is applied to the TF bins that pass the coherence test. Therefore the selected RTF features are supposed to have the same robustness to noise as the proposed DP-RTF features.

\subsubsection{The Model-based EM Source Separation and Localization method (MESSL) \cite{mandel2010}} This method is based on a GMM-like joint model of ILD and IPD distribution. MESSL is a semi-blind method, i.e. the number of speakers on a given analyzed sound sequence is assumed to be known. 
\addnote[messlspe]{1}{We used the implementation provided by the authors.\footnote{https://github.com/mim/messl} The default setup is used for the  parameter initialization and tying scheme, 
namely the GMM weights are initialized using a cross-correlation method while the other parameters are initialized in a non-informative way, and the parameters are not tied at all. 
A pilot comparison was conducted to test the three different configurations: i) default, with ILD but not garbage source, ii) without both ILD and garbage source, and iii) with both ILD and garbage source. 
The third configuration slightly outperformed the other two, thus it was adopted in the following experiments. } 
For the binaural dataset, the set of candidate delays corresponds to the azimuth grid used for the proposed method, and they are computed from the corresponding HRIRs. 
\addnote[mcmessl]{1}{For the multichannel robotic head data, the multichannel MESSL proposed in \cite{mandel2016} is used. The set of candidate delays is uniformly distributed in the possible maximum range. 
Source localization is made by comparing the output multichannel delays and the delay templates corresponding to the azimuth grid used for the proposed method.}

\subsubsection{The Steered-Response Power using the PHAse Transform (SRP-PHAT) \cite{dibiase2001,do2007}} This is a classic one-stage algorithm. The candidate azimuth directions of the proposed method are taken as the steering directions, and the corresponding HRIRs are used as the steering responses.
The number of sources and their locations can be detected by selecting the peaks with steered response power above a threshold. However, the steered response power for different acoustic conditions, such as different number of sources, SNRs, or reverberation times, can significantly vary, which makes the threshold setting difficult. Thence, in the following experiments, we use SRP-PHAT in a semi-blind mode.

\subsection{Results of Experiments with Simulated Data}
In this subsection, we first present an example of result obtained on simulated data to illustrate the behavior of the localization methods, and then we provide more general quantitative results. \addnote[epmle]{1}{We remind that the proposed method is referred to as EP-MLE. }

\subsubsection{An Example of Sound Source Localization}

\begin{figure}[t]
\centering
\subfloat[EP-MLE, DRR=0.5dB]{\includegraphics[width=0.45\columnwidth]{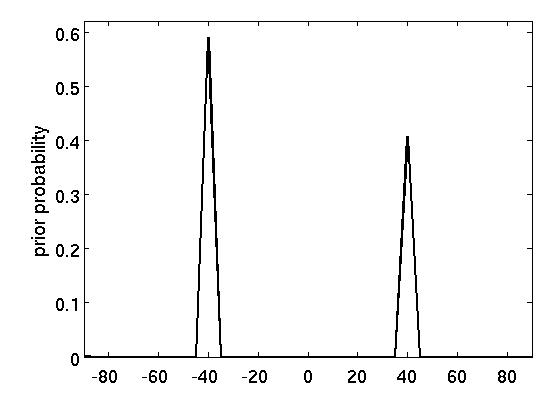}}
\subfloat[EP-MLE, DRR=-5.5dB]{\includegraphics[width=0.45\columnwidth]{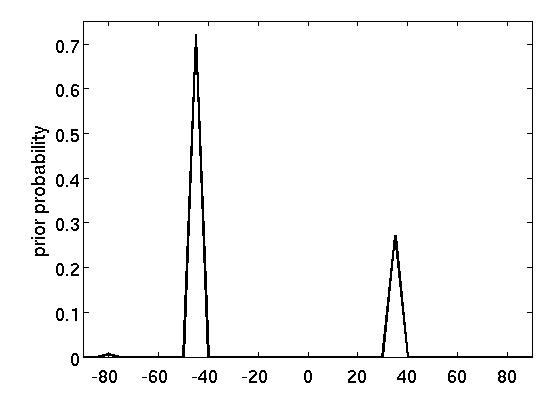}} \\
\subfloat[Basic-CGMM, DRR=0.5dB]{\includegraphics[width=0.45\columnwidth]{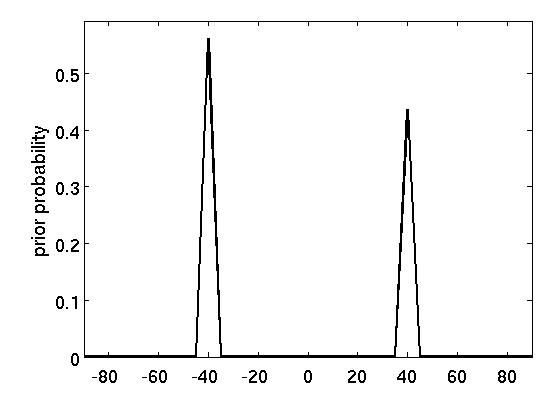}}
\subfloat[Basic-CGMM, DRR=-5.5dB]{\includegraphics[width=0.45\columnwidth]{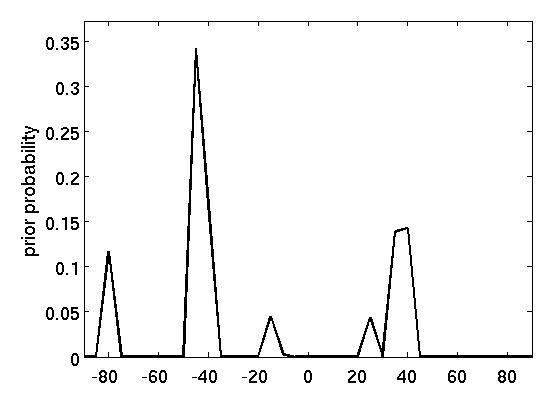}} \\
\subfloat[RTF-CT-CGMM, DRR=0.5dB]{\includegraphics[width=0.45\columnwidth]{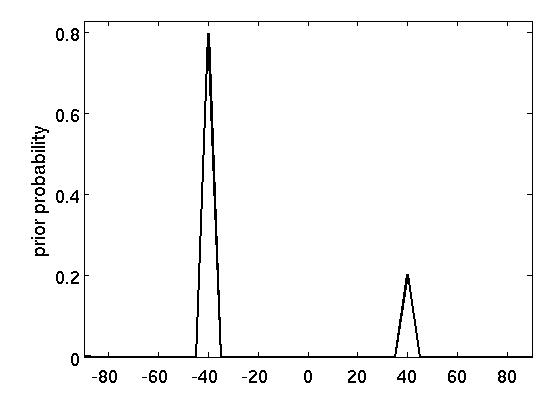}}
\subfloat[RTF-CT-CGMM, DRR=-5.5dB]{\includegraphics[width=0.45\columnwidth]{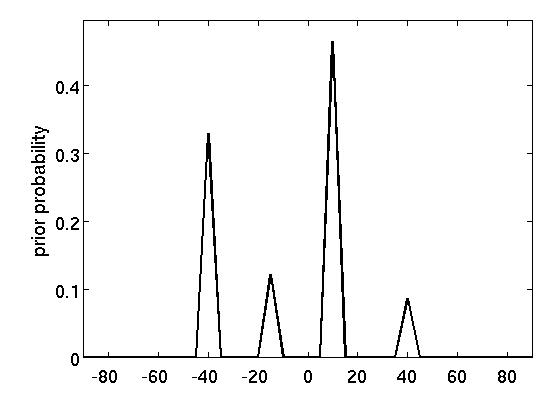}} \\
\subfloat[MESSL, DRR=0.5dB]{\includegraphics[width=0.45\columnwidth]{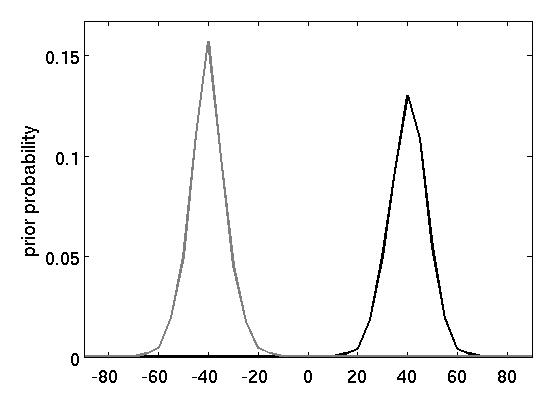}}
\subfloat[MESSL, DRR=-5.5dB]{\includegraphics[width=0.45\columnwidth]{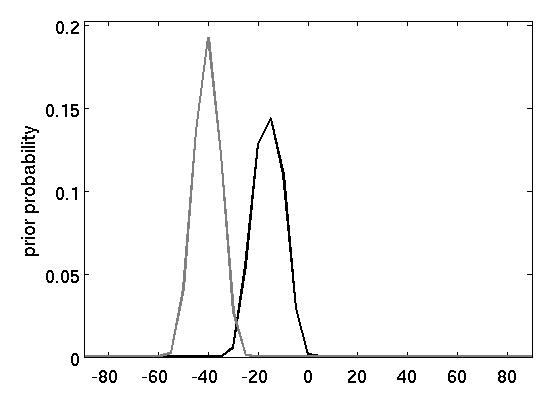}} \\
\subfloat[SRP-PHAT, DRR=0.5dB]{\includegraphics[width=0.45\columnwidth]{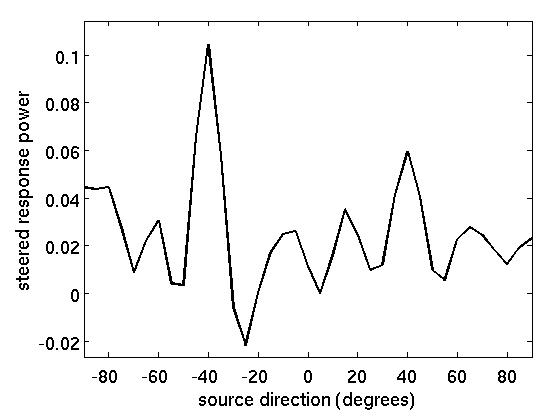}} 
\subfloat[SRP-PHAT, DRR=-5.5dB]{\includegraphics[width=0.45\columnwidth]{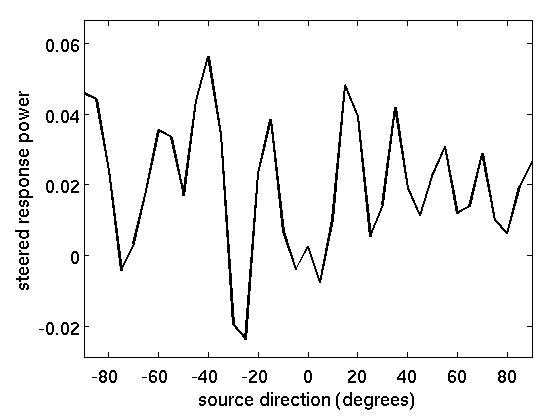}} 
\caption{\small{An example of source localization obtained with the proposed method and with the four baseline methods. Two speakers located at azimuths $-40^\circ$ and $40^\circ$. SNR is $30$~dB.}}
\label{fig:localization}
\vspace{-3mm}
\end{figure}

\begin{table*}
\caption{\small{Semi-blind localization results for simulation data under various acoustic conditions. The lowest outlier rate among five methods for each acoustic condition is shown in \textbf{bold}.}}
\centering\small
\begin{tabular}{|c| c  c | c  c   | c   c | c   c | c   c |c  c|}
\hline         
	    &SNR     &DRR          & \multicolumn{2}{c|}{EP-MLE (prop.)}        & \multicolumn{2}{c|}{Basic-CGMM}      & \multicolumn{2}{c|}{RTF-CT-CGMM}     & \multicolumn{2}{c|}{MESSL \cite{mandel2010}}   & \multicolumn{2}{c|}{SRP-PHAT \cite{do2007}}    \\ 
	    &(dB)    &(dB)         &Out(\%) & MAE($^\circ$)               & Out(\%)  & MAE($^\circ$)            & Out(\%)   & MAE($^\circ$)               & Out(\%)  & MAE($^\circ$)     & Out(\%)    & MAE($^\circ$)       \\ \hline 
	    &30      & 0.5         &0.9     & 0.15               	  & \textbf{0.2}      & 0.18           	        & 5.6       & 1.91                        & 0.4      & 0.14              & 2.0        & 0.42                \\ 
Two 	    &30      & -5.5        &\textbf{2.3}     & 2.06                	  & 3.6      & 2.03           		& 26.4      & 4.71                        & 27.7     & 2.06        	 & 34.8       & 2.81                \\
speakers     &5       & 0.5         &6.2     & 1.94                	  & \textbf{5.4}      & 1.94           		& 11.5      & 4.53                        & 17.4     & 2.75        	 & 6.1        & 1.75                \\                             
	    &5       & -5.5        &\textbf{15.1}    & 5.12                	  & 18.9     & 5.05           		& 30.1      & 6.31                        & 36.8     & 5.13        	 & 35.7       & 4.30                \\ \hline
	    &30      & 0.5         &3.4     & 0.58                	  & \textbf{1.5}      & 0.64           		& 15.5      & 2.76                        & 2.1      & 0.46        	 & 5.5        & 0.98                \\ 
 Three      &30      & -5.5        &\textbf{12.9}    & 2.93                	  & 16.1     & 2.91           		& 29.9      & 5.54                        & 35.7     & 2.55        	 & 35.6       & 3.18                \\
 speakers    &5       & 0.5         &18.7    & 3.05                	  & 17.2     & 3.08           		& 19.7      & 5.29                                & 24.1     & 3.29        	 & \textbf{13.4}       & 2.52                \\                             
	    &5       & -5.5        &\textbf{23.1}    & 5.53                	  & 25.6     & 5.49           		& 33.7      & 6.64                        & 37.5     & 5.10        	 & 34.7       & 5.03                \\ \hline
\end{tabular}
\vspace{-0mm}
\label{tab:results}
\end{table*}

\begin{table*}
\caption{\small{Blind localization results for simulation data under various acoustic conditions. The lowest MD and FA among three methods for each acoustic condition are shown in \textbf{bold}.}}
\centering\small
\begin{tabular}{|c| c  c | c  c  c | c  c  c | c  c  c |}
\hline         
	    &SNR        &DRR       & \multicolumn{3}{c|}{EP-MLE (prop.)}       & \multicolumn{3}{c|}{Basic-CGMM}    & \multicolumn{3}{c|}{RTF-CT-CGMM}     \\ 
	    &(dB)     &(dB)        &MD(\%) & FA(\%) & MAE($^\circ$)      & MD(\%)& FA(\%)   & MAE($^\circ$)   & MD(\%) & FA(\%)    & MAE($^\circ$)      \\ \hline
	    &30      & 0.5         &6.2    & \textbf{0}      & 0.15               & \textbf{1.8}   & 1.5      & 0.17            & 11.9   & 12.0      & 1.81               \\ 
Two 	    &30      & -5.5        &\textbf{4.1}    & \textbf{6.6}    & 1.75               & 9.1   & 6.7      & 1.75            & 28.3   & 37.7      & 5.03               \\
speakers     &5       & 0.5         &\textbf{13.4}   & \textbf{0.3}    & 1.68               & 17.4  & 1.2      & 1.70            & 14.4   & 17.3      & 4.45               \\                             
	    &5       & -5.5        &\textbf{16.1}   & \textbf{15.7}   & 4.88               & 21.7  & 17.3     & 4.79            & 30.5   & 37.4      & 6.68               \\ \hline
	    &30      & 0.5         &\textbf{17.9}   & \textbf{0.2}    & 0.53               & 18.5  & 0.5      & 0.48            & 27.1   & 10.0      & 2.57              	\\ 
 Three      &30      & -5.5        &\textbf{19.9}   & \textbf{9.3}    & 2.61               & 22.4  & 12.4     & 2.74            & 40.6   & 20.7      & 5.30              	\\
 speakers    &5       & 0.5         &\textbf{29.2}   & \textbf{2.3}    & 2.80               & 31.2  & 4.6      & 2.83            & 29.7   & 15.1      & 5.38               \\                             
	    &5       & -5.5        &\textbf{31.9}   & \textbf{15.3}   & 5.41               & 33.8  & 18.3     & 5.85            & 42.2   & 22.1      & 6.78              	\\ \hline
\end{tabular}
\vspace{0mm}
\label{tab:resultb}
\end{table*}

Fig.~\ref{fig:localization} shows a source localization example obtained with the proposed method and with the baseline methods. 
For DRR~=~$0.5$~dB (left column), all methods (except for SRP-PHAT) have two (and only two) prominent peaks at the correct source azimuths. The SRP-PHAT profile is more cluttered than the other profiles but the two highest peaks are nevertheless at the correct source azimuth. The results for the proposed EP-MLE and the Basic-CGMM method are quite similar, hence the entropy penalty has no significant influence in these conditions. It can be seen that EP-MLE, Basic-CGMM, and RTF-CT-CGMM (hence all CGMM-based methods) have narrower peaks than MESSL. 
The reason is that, after spectral subtraction, the proposed DP-RTF features and the RTF features are less noisy than the ILD/IPD used in MESSL.

For DRR~=~$-5.5$~dB (right column), the source at $-40^\circ$ still has a prominent peak for the first four methods (though the maximum of the peak is slightly shifted at $-45^\circ$ for EP-MLE and Basic-CGMM). Even the SRP-PHAT profile, though made very hectic by the intense reverberations, keeps its maximum at $-40^\circ$. However, the source at $40^\circ$ does not have a very large peak for RTF-CT-CGMM, whereas there is a much higher peak at $10^\circ$. One possible reason for this is that a high amount of reverberations decreases the number of TF bins dominated by the direct-path propagation of a single source, hence a lower number of TF bins can be selected by the coherence test. In addition, an improper threshold can make the detected TF bins involve reflections. MESSL fails to detect the source at $40^\circ$ as well: there are still two prominent peaks but the second one is clearly misslocated at $-15^\circ$. The reason for this is that the ILD/IPD features are heavily contaminated by strong reverberations. Finally, the very hectic profile of SRP-PHAT does not allow the detection of the second source.
In contrast, it can be seen that the proposed EP-MLE and Basic-CGMM methods provide second-prominent peaks at the correct source location (actually at $35^\circ$ for EP-MLE).
This again shows that, compared with the MTF-based RTF feature, the proposed DP-RTF feature is more reliable for multi-source localization in highly reverberant environments.  
In addition to the peaks at the correct azimuths, there are also a few other spurious peaks in the case of the Basic-CGMM method. The use of the entropy penalty in EP-MLE successfully suppresses the spurious peaks and strengthen the true peaks.
This illustrates well the sparsity-enforcing property of the entropy penalty term. For the semi-blind configuration, correct localization is obtained by both EP-MLE and Basic-CGMM, in this example. But in the blind configuration, the selection threshold is very difficult to set automatically for the Basic-CGMM method, due to amplitude similarity of the correct peak at $40^\circ$ and of the spurious peak at $-80^\circ$. This may easily lead to either miss detection or false alarm. In contrast, the EP-MLE method enables a large range of threshold values that lead to correct detection in this example. 
Note that there is a larger risk of errors for Basic-CGMM even in the semi-blind configuration: a slightly larger spurious peak at $-80^\circ$ would lead to a wrong localization.

\subsubsection{Semi-blind Localization Results}
Table~\ref{tab:results} shows the semi-blind localization results obtained for various acoustic conditions, averaged over the $500$ above-mentioned test mixtures.
We first compare the two-speaker localization results of the proposed method with the results of MESSL and SRP-PHAT. For SNR $=30$~dB and DRR $=0.5$~dB, all three methods achieve satisfactory and comparable performance. When only the DRR decreases (to $-5.5$ dB), the outlier rate of MESSL and SRP-PHAT dramatically increases, whereas the outlier rate of the proposed method increases only slightly.
This indicates that the ILD/ITD features and the steered response power are less robust to reverberations than the proposed DP-RTF features. 
For MESSL, the garbage source is not able to collect the colored interfering features caused by the intense reverberations. When only the SNR decreases (to $5$ dB), the performance measures of all the three methods degrade, as expected. For EP-MLE, the noise residual after spectral subtraction is larger for the low SNR case. Moreover, more frames with low speech power are highly corrupted by noise, which decreases the number of valid TF bins used for DP-RTF estimation. For MESSL, the estimated ILD/ITD features are severely corrupted by the noise, especially by the directional (spatially correlated) noise. In addition, the ILD/ITD extracted from the TF bins dominated by the directional noise will lead to a spurious peak in the noise direction. For these reasons, MESSL performs the worst out of the three methods (at SNR = $5$~dB). 
For SRP-PHAT, the directional noise also contaminates the steered response power, possibly leading to a spurious peak. SRP-PHAT outperforms MESSL, and is comparable with the proposed method, possibly due to the efficiency of PHAT weight.
When both SNR and DRR are low ($5$ dB and $-5.5$ dB, respectively), the proposed method prominently outperforms the two other methods in terms of outlier rate. 

We then analyze the three-speaker localization results. Compared to the two-speaker case, the localization performances of all methods degrade, as expected. Indeed, the WDO assumption is less valid as the number of sources increases, i.e. the number of TF regions that are dominated by a single source decreases. For the proposed method, this leads to a lower number of DP-RTF observations and worse localization performance. For MESSL, this leads to estimated ILD/ITD features that are less reliable, which also leads to a worse localization performance. For SRP-PHAT, the multiple sources can be mutually considered as noise signals, so more sources will make the steered response power of the actual source directions less significant.
Overall, the proposed method globally outperforms MESSL and SRP-PHAT, except for DRR $=0.5$~dB and SNR $=5$~dB, for which SRP-PHAT performs the best.


One can see from Table~\ref{tab:results} that the proposed method outperforms the RTF-CT-CGMM method for all acoustic conditions.
Therefore, it is confirmed that the proposed CTF-based DP-RTF feature combined with the proposed consistency test provides more reliable features than the usual MTF-based RTF combined with the coherence test.
As for Basic-CGMM, the DP-RTF estimation error for DRR~$=-5.5$~dB will lead to noticeable spurious peaks, as was illustrated in Fig.~\ref{fig:localization}. By suppressing the spurious peaks and/or strengthening the correct peaks, thanks to the entropy penalty, the proposed EP-MLE method achieves a significantly smaller outlier rate than Basic-CGMM, for a similar MAE. However, for DRR~$=0.5$~dB, there are much less spurious peaks, or they are much lower than the correct peaks. Thence, the proposed entropy penalty term is here less helpful compared with the low DRR case. 

\subsubsection{Blind Localization Results}
Table~\ref{tab:resultb} shows the blind localization results for the EP-MLE, Basic-CGMM and RTF-CT-CGMM methods. 
It can be seen that, for all three methods, the average of the MD rate and FA rate is generally larger than the outlier rate in the semi-blind configuration, which verifies that the blind configuration is more difficult than the semi-blind one.  
Also, for all methods and in a very general manner, both MD and FA increase when either the SNR or the DRR decreases, and when the number of speaker goes from two to three, which was expected. For the proposed EP-MLE method in particular, a larger DP-RTF estimation error is caused by more intense reverberations, which lead to more spurious peaks and peak shifts. 
\addnote[flack]{1}{For a given DRR, MD increases with the decrease of the SNR or with the increase of the number of speakers, since, as mentioned above, the method may suffer from a lack of sufficient number of DP-RTF observations. } 
When the acoustic conditions get worse in terms of SNR or DRR, MAE increases due to the larger DP-RTF estimation error.  

In general, MD, FA and MAE are considerably smaller for the proposed EP-MLE method (and for Basic-CGMM) compared to the RTF-CT-CGMM method, which is consistent with the results obtained for the semi-blind configuration. 
Unlike the semi-blind configuration, it can be seen that MD and FA are both smaller for EP-MLE than for Basic-CGMM, while the MAE are comparable, for almost all acoustic conditions (all except for MD at SNR~$=30$~dB, DRR~$=0.5$~dB, 2 speakers). 
This confirms the importance of the penalty term in the blind configuration. The semi-blind configuration inherently limits the FA score, and at the same time it can ``force'' the detection of low peaks, ensuring correct MD scores. In contrast, the setting of the threshold in the blind configuration favours either the MD or the FA. Therefore, in the blind configuration, it is more crucial to reduce the spurious peaks and enhance the correct peaks to facilitate the thresholding operation, which is exactly what is done by the entropy penalty term.   
By reducing the entropy to a proper extent, usually, the CGMM component weights corresponding to interfering directions are significantly decreased, while
the weights of the true source directions are enhanced. As a result, MD and FA are both decreased by the entropy penalty term.  

\subsection{Results of Experiments with NAO Head Data}
\begin{table*}[t]
\caption{\small{Semi-blind localization results for NAO data under various acoustic conditions. Here MC-MESSL denotes multichannel MESSL method. The lowest MD and FA among three blind methods for each acoustic condition are shown in \textbf{bold}.}}
\centering\small
\begin{tabular}{|c| c | c    c | c    c | c  c| c c |c c |}
\hline         
	    &robot-to-source          & \multicolumn{2}{c|}{EP-MLE (prop.)}       & \multicolumn{2}{c|}{Basic-CGMM}           &\multicolumn{2}{c|}{RTF-CT-CGMM}  & \multicolumn{2}{c|}{MC-MESSL \cite{mandel2016}}  &\multicolumn{2}{c|}{SRP-PHAT \cite{do2007}}        \\ 
	    &distance                 &Out(\%)     & MAE($^\circ$)      & Out(\%) & MAE($^\circ$)     & Out(\%) & MAE($^\circ$)  & Out(\%) & MAE($^\circ$)   & Out(\%)  &  MAE($^\circ$) \\ \hline
Two	    &1.5 m                    & \textbf{8.5}   & 3.71               & 12.5   &  3.86              & 28.0   & 3.84        & 42.7 & 4.23      & 39.8 & 3.14        \\ 
speakers    &2.5 m                    & \textbf{15.3}  & 4.93               & 21.0   & 5.20              & 24.5   & 5.81        & 44.8 & 4.65     & 36.3 & 4.68  \\ \hline  
Three 	    &1.5 m                    & \textbf{14.5}  & 5.21               & 17.3   & 4.46               & 34.7   & 4.66       & 46.1 & 4.77       & 44.2 & 3.58      \\ 
speakers     &2.5 m                   & \textbf{18.7}  & 5.35               & 22.3   & 5.59              & 22.3   & 5.90        & 52.4 & 5.89     & 47.5 & 5.27    \\ \hline     
\end{tabular}
\label{tab:resultsNAOs}
\end{table*}

\begin{table*}[t]
\caption{\small{Blind localization results for NAO data under various acoustic conditions. The lowest MD and FA among three blind methods for each acoustic condition are shown in \textbf{bold}.}}
\centering\small
\begin{tabular}{|c| c | c  c  c | c  c  c | c c c|c c |}
\hline         
	    &robot-to-source          & \multicolumn{3}{c|}{EP-MLE (prop.)}       & \multicolumn{3}{c|}{Basic-CGMM}           &\multicolumn{3}{c|}{RTF-CT-CGMM}       \\ 
	    &distance                 &MD(\%) & FA(\%) & MAE($^\circ$)      & MD(\%) & FA(\%) & MAE($^\circ$)     & MD(\%) & FA(\%) & MAE($^\circ$)     \\ \hline
Two	    &1.5 m                    & \textbf{8.0}   & 14.3   & 3.79               & 15.5   & \textbf{13.5}   & 3.80              & 33.5   & 22.0   & 4.36                     \\ 
speakers    &2.5 m                    & \textbf{12.8}  & \textbf{18.0}   & 5.60               & 14.0   & 30.5   & 5.38              & 25.0   & 20.5   & 5.06               \\ \hline  
Three 	    &1.5 m                    & \textbf{17.8}  & 15.3   & 4.24               & 24.8   & \textbf{15.2}   & 4.17              & 46.8   & 21.7   & 4.23                  \\ 
speakers     &2.5 m                   & \textbf{20.8}  & 17.7   & 5.37               & 21.8   & 24.3   & 5.45              & 37.2   & \textbf{10.7}   & 5.23                \\ \hline     
\end{tabular}
\label{tab:resultsNAOb}
\end{table*}

Table~\ref{tab:resultsNAOs} and Table~\ref{tab:resultsNAOb} show the source localization results obtained with NAO head data, in the semi-blind and blind configuration, respectively.   
From Table~\ref{tab:resultsNAOs}, it can be seen that the proposed method achieves the lowest outlier rate for all conditions, which verifies the effectiveness of the proposed entropy penalty and DP-RTF feature in such realistic scenarios.
\addnote[messl]{1}{Overall, the multichannel MESSL method performs the worst. On the one hand, the spatially correlated noise and intense reverberation influence the performance as for the binaural case. 
On the other hand, the multichannel MESSL coordinates the microphone pairs through TF masks, rather than the usually used microphone calibration information. The time delay of each microphone pair is estimated using the results of source separation.
This is advantageous for source separation that does not require information on microphone configuration. 
However, the known microphone configuration is necessary for source localization to specify the spatial relation between physical positions, namely to calibrate the time delays. 
In these experiments, the microphone calibration information is only used for source localization by comparing the calibrated time delays and the estimated time delays of MESSL, which leads to unsatisfactory localization performance. }
SRP-PHAT also has a high outlier rate due to the spatially correlated noise and real world reverberations. 

In general, the performance measures reported in Table~\ref{tab:resultsNAOb} are consistent with the  results obtained on the simulated data. 
Compared to Basic-CGMM, EP-MLE has smaller MD under all conditions, smaller FA under two conditions out of four (and for the other two conditions, the FA values for both methods are very close), and a comparable MAE. 
Also, the proposed method significantly outperforms the RTF-CT-CGMM method, since, again, the quantity and the quality of the observations are both higher for the proposed DP-RTF features than the RTF features based on the coherence test.

\section{Conclusion}\label{sec:conclusion}
In this paper, we presented a method for multiple-source localization in reverberant and noisy environments. The method is based on the model of \cite{dorfan2015} with the following original contributions: 
(i) the use of an entropy-based penalty term which enforces sparsity for the estimation of the model parameters, implemented via a convex-concave optimization procedure that is more efficient than an EM algorithm, (ii) the use of DP-RTF features, providing localization that is robust to both noise (thanks to the inter-frame power spectral subtraction) and reverberations, and (iii) the proposed consistency test algorithm that ensures that DP-RTF features are estimated from frame regions associated to a single active speaker, thus making possible to use these features for multiple-speaker localization.
Overall, experiments conducted on both simulated and real-world data show that (i) the proposed DP-RTF features are more reliable than classical MTF-based features, for instance RTF features, (ii) the proposed CGMM model with DP-RTF features provides a better source localization compared to three baseline methods (RTF-based, MESSL, SRP-PHAT) in a semi-blind configuration, and (ii) the entropy penalty term used in the proposed localization technique makes it able to better localize the sources compared to the basic version of the same method (i.e. without the entropy penalty term); this is especially true in a blind configuration where the proposed method is efficient in \emph{jointly counting and localizing the sources}. 
The experiments showed that the entropy-based penalty significantly improves the localization performance in terms of missed detections and false alarms.   

In this study, the entropy-based penalty weighting coefficient $\gamma$ was set to an empirical fixed value leading to good overall performance for all tested conditions. In future work, a principled setting of $\gamma$ could be investigated, considering the noise level of the DP-RTF observations.
Also, the DP-RTF features are more robust than MTF-based features at the cost of the need for more reliable data. 
An improved DP-RTF estimation process requiring less data will be investigated in the near future.



\end{document}